\documentclass[aps,prd,twocolumn,showpacs,10pt,superscriptaddress,preprintnumbers,nofootinbib,floatfix]{revtex4-2}
\usepackage{amsmath,amssymb,mathtools,bm} \allowdisplaybreaks[1]
\usepackage{graphicx} \graphicspath{{figs/}}
\usepackage{placeins}
\usepackage{physics, esint}
\usepackage{xcolor}
\usepackage{silence}
\usepackage[colorlinks, hypertexnames=false, linkcolor=blue!80!black, citecolor=green!50!black, urlcolor=magenta]{hyperref}
\usepackage{adjustbox}
\usepackage{tikz}
\usetikzlibrary{arrows.meta}
\usepackage{enumitem}
\usepackage{soul}
\usepackage{multirow}
\usepackage{orcidlink}
\usepackage[normalem]{ulem}
\newcommand{\fig}[1]{Fig.~\ref{#1}}

\newcommand{\eq}[1]{Eq.~\eqref{#1}}
\newcommand{\eqs}[2]{Eqs.~\eqref{#1} and~\eqref{#2}}

\newcommand{\refcite}[1]{Ref.~\cite{#1}}
\newcommand{\refs}[1]{Refs.~\cite{#1}}

\newcommand{\pp}[1]{\!\left(#1\right)}
\newcommand{\bb}[1]{\!\left[#1\right]}
\newcommand{\cc}[1]{\!\left\{#1\right\}}
\newcommand{\vv}[1]{\!\left\langle #1 \right\rangle}
\newcommand{\bigpp}[1]{\big(#1\big)}
\newcommand{\bigbb}[1]{\big[#1\big]}

\newcommand{\biggpp}[1]{\bigg(#1\bigg)}
\newcommand{\biggbb}[1]{\bigg[#1\bigg]}
\newcommand{\biggcc}[1]{\bigg\{#1\bigg\}}

\newcommand{\beq}[1][]{\begin{equation}\label{#1}}
\newcommand{\eeq}{\end{equation}}
\newcommand{\bse}[1][]{\begin{subequations}\label{#1}}
\newcommand{\ese}{\end{subequations}}
\newcommand{\nn}{\nonumber}

\newcommand{\GeV}{{\rm GeV}}

\newcommand{\qf}{\Lambda_F}

\newcommand{\unpfec}{\mathcal{D}}
\newcommand{\colfec}{\mathcal{H}}

\newcommand{\qfcolcut}{\lambda_{\colfec}}
\newcommand{\E}{\mathcal{E}}

\begin{document}

\title{A universal probe of QCD spin physics from fragmentation energy correlator}

\author{Qing-Hong Cao\,\orcidlink{0000-0003-0033-2665}}
\affiliation{School of Physics, Peking University, Beijing 100871, China}
\affiliation{School of Physics, Zhengzhou University, Zhengzhou 450001, China}
\affiliation{Center for High Energy Physics, Peking University, Beijing 100871, China}

\author{Xiaohui Liu\,\orcidlink{0000-0002-7701-1205}}
\affiliation{Center of Advanced Quantum Studies, School of Physics and Astronomy, Beijing Normal University, Beijing 100875, China}
\affiliation{Key Laboratory of Multi-scale Spin Physics, Ministry of Education, Beijing Normal University, Beijing 100875, China}

\author{Zhite Yu\,\orcidlink{0000-0003-1503-5364}}
\affiliation{High Energy Theory Group, Physics Department, Brookhaven National Laboratory, Upton, New York 11973, USA}

\author{C.-P. Yuan\,\orcidlink{0000-0003-3988-5048}}
\affiliation{Department of Physics and Astronomy, Michigan State University, East Lansing, Michigan 48824, USA}

\author{Feng Yuan\,\orcidlink{0000-0001-9979-3853}}
\affiliation{Nuclear Science Division, Lawrence Berkeley National Laboratory, Berkeley, California 94720, USA}

\author{Shu-Tao Zhang\,\orcidlink{0009-0009-5269-2364}}
\affiliation{School of Physics, Peking University, Beijing 100871, China}

\author{Hua Xing Zhu\,\orcidlink{0000-0002-7129-6748}}
\affiliation{School of Physics, Peking University, Beijing 100871, China}
\affiliation{Center for High Energy Physics, Peking University, Beijing 100871, China}

\date{\today}

\begin{abstract}
We study the fragmentation energy correlator (FEC), 
which combines the nonperturbative fragmentation function with the infrared-safe energy correlator,
as a new probe of the three-dimensional dynamics of QCD hadronization.
We demonstrate, both theoretically and numerically, that FEC applies universally to inclusive hadron production in
lepton-lepton, lepton-hadron, and hadron-hadron collisions, without requiring jet algorithms or simultaneous measurements of other heavy objects.
Through the nontrivial transverse-spin dependence,
FEC provides a sensitive probe of the chiral-symmetry-breaking scale in QCD via its variation with the energy-correlator scale.
\end{abstract}

\makeatletter
\@author@finish
\let\sm@title\@title
\let\sm@authors\@AAC@list
\let\sm@affiliations\@AFF@list
\makeatother
\maketitle

\emph{Introduction.}---
Energy correlations are observables at collider experiments 
that measure the angular correlation of energy flows in final states. 
They are first introduced as infrared safe observables 
to test the perturbative regime of QCD~\cite{Basham:1978bw, Basham:1978zq, Basham:1977iq,Basham:1979gh},
and have recently resurged and evolved into precision tools 
applied to a wide range of phenomena; see \refcite{Moult:2025nhu} for a recent review.
The cleanest application of these is to the $e^+e^-$ annihilation,
where the QCD interactions only concern the final states,
allowing the energy correlations to be described by vacuum expectation values
of energy flow operators $\E(\bm{n})$~\cite{Sterman:1975xv, Sveshnikov:1995vi, Korchemsky:1999kt,Belitsky:2001ij,Belitsky:2013xxa} 
and their track generalization~\cite{Chang:2013rca,Chang:2013iba,Jaarsma:2023ell,Chen:2022pdu,Chen:2022muj,Li:2021zcf,Lee:2023xzv,Lee:2026hub},
which effectively probe the vacuum at a given center-of-mass (c.m.) energy scale $Q$.
The latter are further related to conformal field theory~\cite{Hofman:2008ar}
and enable powerful methods of light-ray operator analysis~\cite{Kravchuk:2018htv, Kologlu:2019mfz, Chang:2020qpj, Chen:2022jhb, Caron-Huot:2022eqs, Chen:2020adz, Chen:2021gdk, Chen:2023zzh}.

Phenomenologically, energy correlators provide an alternative way to 
study jets in hadronic final states without the need for jet algorithms~\cite{Dixon:2019uzg, Chen:2020vvp}.
This makes it easier to compare theory and experiment at ultrahigh precision~\cite{Lee:2022uwt,Chen:2023zlx,Lee:2026zyl}.
For example, the study of two-point energy correlator (EEC) has produced a full picture of the transition
from the perturbative collinear scaling governed by parton branching and light-ray operator evolution
to the confinement regime~\cite{Liu:2024lxy,Barata:2024wsu,Lee:2025okn,Guo:2025zwb,Chang:2025kgq,Kang:2025zto,Herrmann:2025fqy,Guo:2026bhd}
as the angle $\theta$ between the two energy flows becomes increasingly small~\cite{Komiske:2022enw,CMS:2024mlf,CMS-PAS-HIN-24-019,Electron-PositronAlliance:2025fhk,ALICE:2024dfl,STAR:2025jut}.
In this process, the spectrum of $Q\theta$ plays the role as a scale scanner from parton radiation to hadron formation.

While the successful application of energy correlators in $e^+e^-$ collisions 
has motivated their extensions to lepton-proton ($ep$) and proton-proton ($pp$) colliders, a nontrivial question arises as to whether hadronic activities in the initial states
factorize from the jet production in the final states.
This greatly affects whether the same vacuum expectation values of energy correlators at $e^+e^-$ collisions
continue to describe the $ep$ and $pp$ events, that is, the universality of energy correlators.
While phenomenological studies suggest this is likely the case~\cite{Liu:2024lxy, Barata:2024wsu,Guo:2025qnz},
a definitive theoretical argument is still lacking.
In particular, when applied to $pp$ collisions, the EEC alone does not define any hard scale,
so typically requires the additional use of a jet algorithm~\cite{CMS:2024mlf,ALICE:2024dfl,STAR:2025jut} to study energy correlators {\it inside jets}, or the tagging of simultaneously produced heavy objects~\cite{CMS-PAS-HIN-24-019, CMS-PAS-SMP-25-009, Generet:2026vto}.

\begin{figure}
    \centering
    \includegraphics[width=0.9\linewidth]{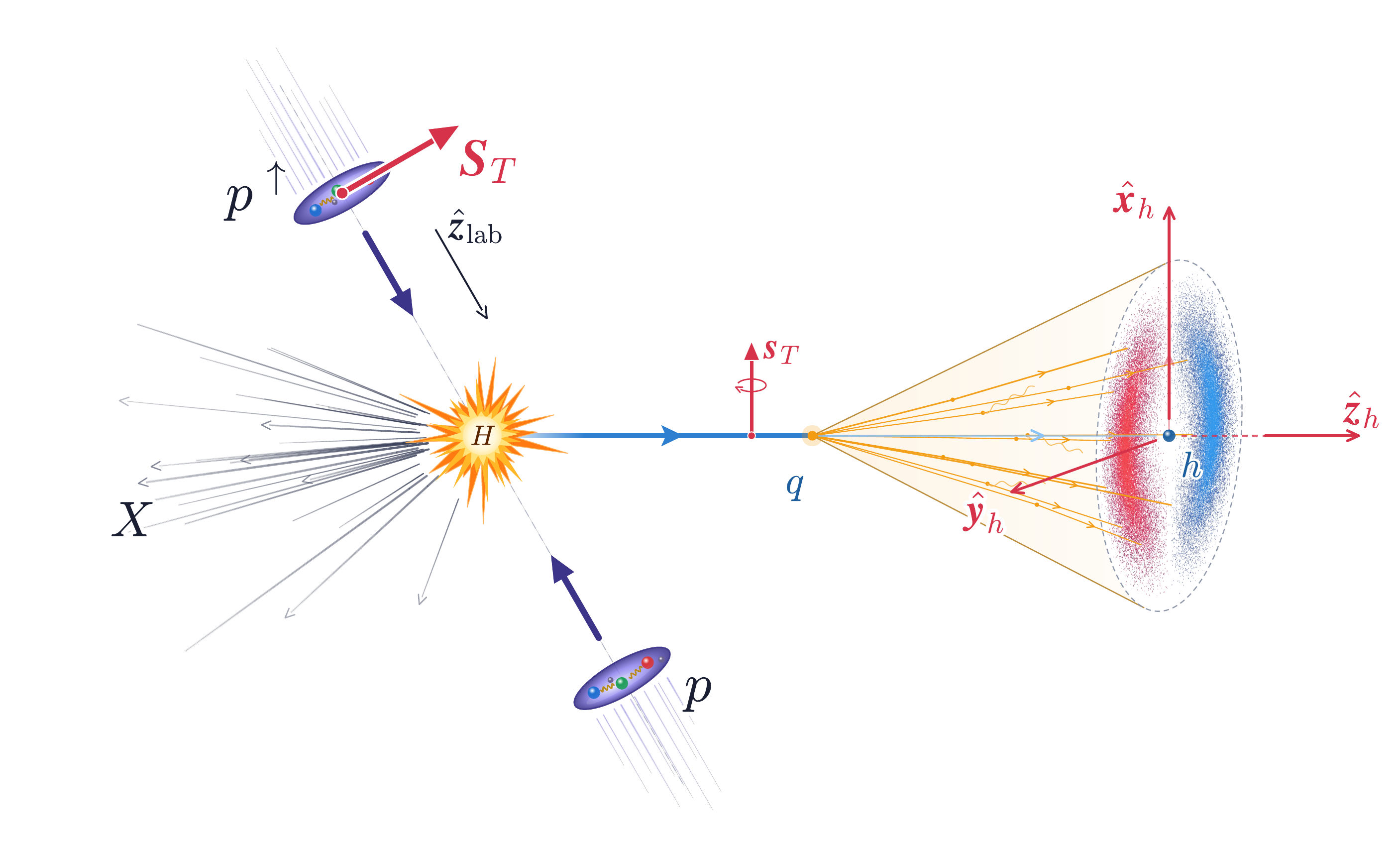}
    \caption{Schematic energy distribution around a tagged hadron
    produced in singly transversely polarized $p^\uparrow p$ collisions. The red (blue) shade indicates enhanced (suppressed) energy deposition in the plane transverse to the hadron momentum.}
    \label{fig:fec-pattern}
\end{figure}

In this Letter, we hope to point out that the {\it fragmentation energy correlator (FEC)}~\cite{Liu:2024kqt,Cao:2025icu, Zhu:2025qkx}
provides the first energy correlator observable that universally applies across $e^+e^-$, $ep$, and $pp$ colliders
without reliance on any jet algorithm or other heavy objects.
As illustrated in \fig{fig:fec-pattern},
FEC quantifies the correlation of a tagged hadron and its nearby energy flow.
The momentum of this hadron defines the hard scale
that is necessary to localize the production process within a short distance.
When the energy flow is close to the hadron, 
which is exactly the region we focus on,
the FEC contains similar collinear dynamics to the hadron fragmentation function (FF),
and thus can be universally factorized in the same way as the latter 
in all inclusive productions.

More importantly, as we will dive further into below, 
the FEC of a transversely polarized quark possesses a nontrivial azimuthal dependence on the energy flow even for an unpolarized hadron,
which offers a 
novel way to probe the spontaneous chiral symmetry breaking in QCD. 
This is similar to the Collins effect
studied in the transverse momentum-dependent (TMD) framework~\cite{Collins:1992kk, Bacchetta:2002tk, Amrath:2005gv}.
However, FECs are collinear-factorized, without relying on a small recoil momentum, and so the associated spin asymmetries are not suppressed
by the Sudakov factor nor distorted by soft gluon radiations~\cite{Boer:2001he, Kang:2015msa, Hatta:2020bgy, Hatta:2021jcd}.
They can therefore be extracted from a global analysis in a similar way to FFs,
but with an additional nonperturbative scale provided by the energy flow that serves as a proxy of the chiral symmetry breaking.

\emph{FEC Definition.}---
The FEC can be viewed as an EEC with one of the energy-flow measurements replaced by a tagged hadron. Based on the original proposal in \refcite{Liu:2024kqt} and 
the improvement in \refcite{Cao:2025icu} to make it boost invariant, a key feature for universality,
it is defined as
\begin{align}
  &\Delta^{\alpha\beta}_{h/q}(z, \qf, \phi)
  = \frac{z}{4N_c} \sum_X
	\int \frac{dw^-}{2\pi p_h^+}\,
	e^{i p_h^+ w^-/z}
	\nn\\
  &\hspace{1em} \times
	\langle 0| \, \psi_{q}^{\alpha}(w^-)\, \E_L(\qf,\phi) \, |h,X\rangle \,
	\langle h,X| \, \bar{\psi}_{q}^{\beta}(0) \, |0\rangle 
	\,,
\label{eq:fec-matrix}
\end{align}
where the spinor indices $\alpha$ and $\beta$ are left open.
In this definition, the observed hadron $h$ is taken to
move along the $\hat z_h$ direction, with respect to which we define the lightfront components $v^\pm \equiv \pp{v^0\pm v^3}/\sqrt 2$.
It carries a fraction $z=p_h^+/k^+$ of the fragmenting quark's longitudinal momentum $k^+$, with $p_h$ the momentum of $h$.
The two future-pointing lightlike Wilson lines are suppressed, and the colors are averaged with $N_c = 3$.

The operator $\E_L(\qf,\phi)$ in \eq{eq:fec-matrix} serves to perform the energy-flow measurement,
defined to act on a final state $|h,X\rangle$ as
\begin{align}
	\E_L(\qf,\phi) \, |h,X\rangle
	= &\sum_{j \in X} p_j^+ \, \delta\bigpp{ \ln\qf - \ln\Lambda_{F,j} } \nn\\
    &\qquad 
    \times \delta(\phi - \phi_j)
		\, |h,X\rangle,
\label{eq:EL-def}
\end{align}
where $\qf$ is a dimensionful variable that resolves the opening angle $\theta$ between the observed energy flux and the tagged hadron through the relation,
\begin{equation}
    \qf = \sqrt{2} p_h^+ \tan(\theta / 2),
\end{equation}
and we denote $\Lambda_{F,j} = \sqrt{2}p_h^+\tan(\theta_j / 2)$ with $\theta_j$ being the angle between particle $j$ and $h$.
The $\E_L$ thus counts the longitudinal momentum $p_j^+$ from particles $j$ in $X$ per unit rapidity
and azimuth $\phi$ around the tagged hadron $h$. 
It is a modification of the original energy flow operator $\E(\bm{n})$ that counts {\it energy} per unit polar angle and can also be related to the energy momentum tensor~\cite{Cao:2025icu}. 
The advantage of using $(p_j^+, \qf, \phi)$ is that the transformation of $\E_L$ under a boost along $\hat z_h$ is just a multiplicative factor, which is canceled by the factor $1 / p_h^+$ in \eq{eq:fec-matrix} to make the FEC boost invariant. 
This improvement is essential for establishing the universality of FECs across experiments conducted at different energies and analyzed in different reference frames, and represents a crucial step toward a global analysis.

The boost-invariant variable $\qf$ is central to our analysis.
In particular, it satisfies $\qf\sim E_h\theta$ at the collinear limit, making its interpretation as a transverse scale explicit.
Together with $\phi$, it provides two-dimensional resolution of the transverse fragmentation structure.
At leading twist, an unpolarized hadron $h$ has only two different FECs,
obtained by tracing the spinor indices $(\alpha, \beta)$ in \eq{eq:fec-matrix} 
with corresponding Dirac projectors,
\bse[eq:fec-spin]\begin{align}
  \frac{1}{2 \qf^2} 
  	{\rm Tr} \!\left[ \gamma^+ \, \Delta_{h/q} \right]
  &= \unpfec_{h/q}(z, \qf^2), 
  \label{eq:fec-def-D}
  \\
  \frac{1}{2 \qf^2}
  	{\rm Tr} \!\left[ \gamma^+\gamma_\perp^i\gamma_5 \, \Delta_{h/q} \right]
  &= \colfec_{h/q}(z, \qf^2) \, \pp{ -\sin\phi, \cos\phi }^i
  \,.
\end{align}\ese
The $\unpfec_{h/q}(z, \qf^2)$ describes unpolarized quark fragmentation,
which gives azimuthally uniform energy spread,
whereas the $\colfec_{h/q}(z, \qf^2)$ arises from a transversely polarized quark,
and is associated with an azimuthally asymmetric energy flow,
whose distribution is related to the quark spin direction $i$ by
$(\hat{z}_h \times \hat e_{\phi} )^i = (- \sin\phi, \cos\phi)^i$,
which is the only form allowed by rotation and parity symmetries,
and where $\hat e_{\phi} = (\cos\phi, \sin\phi)$ is the transverse unit vector of the energy flow.
The pattern is illustrated in Fig.~\ref{fig:fec-pattern}.

The $\colfec_{h/q}(z, \qf^2)$ originates from the interference of opposite quark helicity states.
Similar to the Collins function~\cite{Collins:1992kk,Bacchetta:2001di,Bacchetta:2002tk,Amrath:2005gv,Yuan:2008yv},
it cannot be generated perturbatively and is only nonzero because of the chiral symmetry breaking in QCD.
We will therefore refer to it as Collins FEC.

Similar definitions of FECs hold for antiquarks and gluons,
each having an unpolarized FEC $\unpfec_{h/i}$ and a polarized FEC $\colfec_{h/i}$.
For gluons, however, the $\colfec_{h/g}$ describes linear gluon polarization, 
which is not related to chiral symmetry breaking and 
is associated with $\cos2\phi$ and $\sin2\phi$ asymmetries~\cite{Li:2023gkh, Guo:2024jch, Guo:2024vpe}
rather than $(\cos\phi, \sin\phi)$.
It will not mix with the Collins FECs of quarks and antiquarks,
and will not be delved into in this work.

{We note that FECs directly interpolate between dihadron fragmentation functions (DiFFs) and EECs.
Starting with a DiFF and integrating over one of the hadrons along with its energy weight leads to an FEC, for both unpolarized and polarized cases.
Further summing over $h$ in the FEC with its energy weight will give the EEC jet function~\cite{Kang:2025zto}.
In this sum, however, the Collins FEC automatically vanishes due to the symmetry of EEC.
More details of these relations can be found in the Supplemental Material (SM)~\cite{sm}.
The important result of these relations to our study is that the FEC scale $\qf/z$ has a direct correspondence to the EEC scaling variable $Q\theta$.
Since the latter has been established as a scale scanner across the perturbative to nonperturbative regions~\cite{Liu:2024lxy,Barata:2024wsu,Lee:2025okn,Guo:2025zwb,Chang:2025kgq,Kang:2025zto,Herrmann:2025fqy,Guo:2026bhd}, 
we shall expect the same feature to hold for the $\qf/z$ of FECs.
This observation provides a strong inspiration for our parametrization of FECs below.
In addition to the confinement transition encoded in the EEC, by examining the transition behavior of the Collins FEC in its $\qf$ spectrum, 
we also gain a handle for the QCD chiral symmetry breaking.}

\emph{Modeling of FEC.}---
We concentrate on charged pion production.
As a result of full inclusiveness in the energy flow, 
the FECs satisfy the same charge conjugation and SU(2) isospin relations as the FFs.
In particular, for $u$ and $d$ quarks, which form an isospin doublet,
we have flavor-favored and flavor-unfavored FECs,
$\mathcal{F}_{\pm}
  \equiv \mathcal{F}_{\pi^{\pm}/u}
   = \mathcal{F}_{\pi^{\mp}/\bar u}
   = \mathcal{F}_{\pi^{\mp}/d}
   = \mathcal{F}_{\pi^{\pm}/\bar d}$,
for $\mathcal{F} = \unpfec$ or $\colfec$.

While FECs are nonperturbative and ultimately need to be obtained from global fits of experiments,
the unpolarized FEC can be extracted from {\tt Pythia}~\cite{Bierlich:2022pfr} simulation as a good approximation.
We use the energy flow-weighted inclusive pion production cross section $\Sigma$ at $e^+e^-$ collision to extract the unpolarized quark FECs. At leading order (LO), it takes the factorized form,
\begin{align}
	\frac{d\Sigma_{ee\to hX}}{dz \, d\qf^2}
	&= 2\pi \sigma_0 \sum_{q = u, \bar u, \cdots} e_q^2 \, 
		\mathcal{D}_{h/q}(z, \qf^2; \mu^2 = s/4),
\label{eq:ee-1pI-fec-lo}
\end{align}
which is normalized by 
$\sigma_0 = 4\pi N_c \alpha^2 / (3s)$
and where $h = \pi^{\pm}$ and the sum over $q$ runs over quarks and antiquarks,
with $e_q$ the electric charge and $\alpha$ the fine structure constant.
The gluon channel can be similarly extracted from the simulated events of
$e^+e^- (\to \Phi) \to gg$ mediated by a 
fictitious Higgs-like scalar boson $\Phi$.

\begin{figure}[htbp]
    \centering
	\includegraphics[width=\columnwidth]{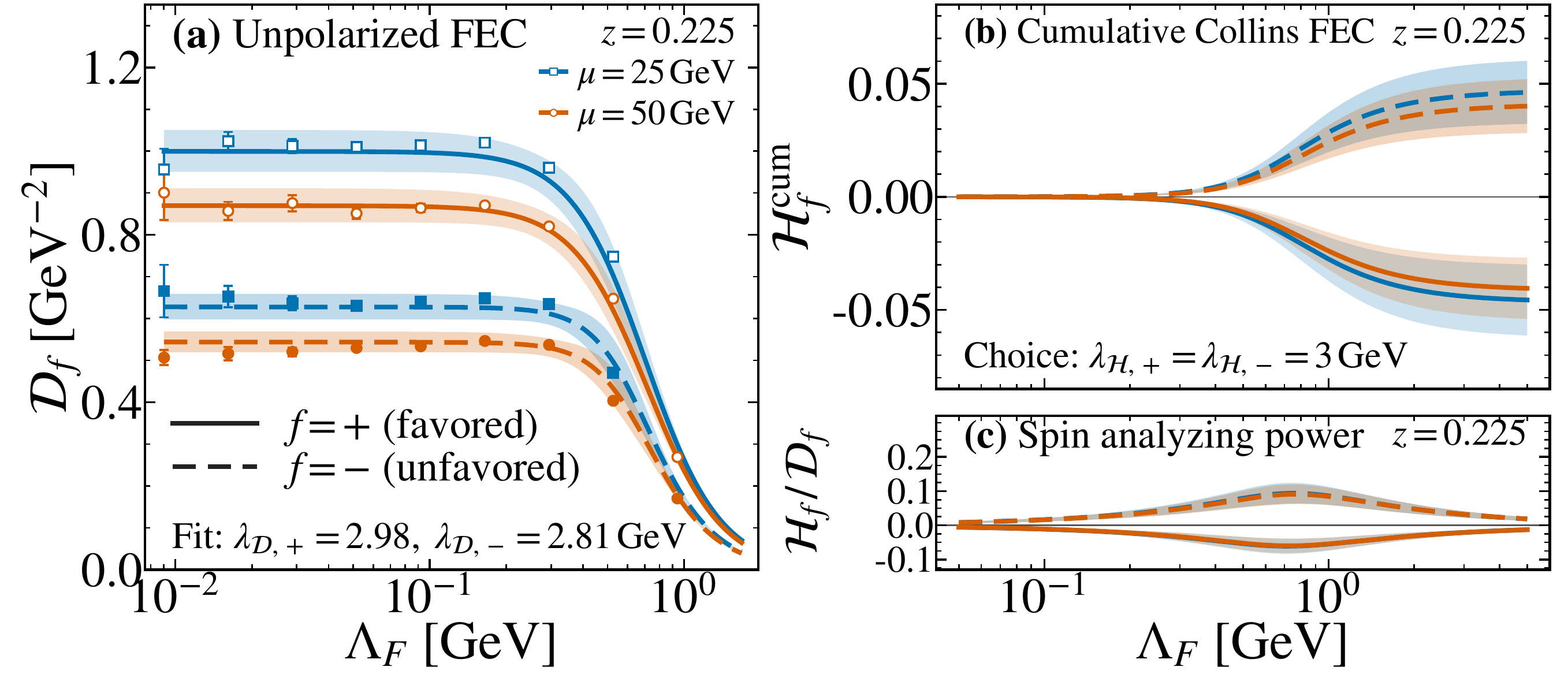}
	\caption{Models of (a) unpolarized FEC, $\unpfec_{\pm}(z,\qf^{2})$, and (b) Collins FEC, $\colfec_{\pm}(z,\qf^{2})$, shown for its cumulative $\qf^{2}$ distribution,
    $\colfec_{\pm}^{\mathrm{cum}}(z,\qf^{2})
    \equiv \int_0^{\qf^{2}} d\lambda^{2}\,
    \colfec_{\pm}(z,\lambda^{2})$,
    along with (c) the spin analyzing power $\colfec_{\pm}(z,\qf^{2}) / \unpfec_{\pm}(z,\qf^{2})$,
    at $z=0.225$.
    In (a), the markers represent the FECs extracted from
    {\tt Pythia} using \eq{eq:ee-1pI-fec-lo};
    the curves and bands show the fitted model and its uncertainty
    at the indicated values of $\lambda_{\unpfec,\pm}$.
    The bands in (b) reflect the {\tt JAM3D-22} Collins function
    uncertainties propagated from $\mu_0=5~\GeV$
    through LO evolution.}
\label{fig:fec-dglap}
\end{figure}

To model the extracted unpolarized FEC, we parametrize the nonperturbative function at the scale $\mu_0 = 5~\GeV$,
the typical energy scale of the data, as
\begin{align}
	&\unpfec_{f}(z, \qf^2; \mu_0)
	= N_{\unpfec, f}(z)\nn\\
	&\hspace{5em} \times
	\biggbb{ \,
		1 + \biggpp{ \frac{\qf}{z \lambda_{\unpfec, f} } }^{\kappa_{\unpfec, f}(z)}
	\, }^{-\frac{2 + \gamma_{\unpfec, f}(z)}{\kappa_{\unpfec, f}(z)}},
\label{eq:fec-param-D}
\end{align}
and evolve it to other scales {by the DGLAP equation~\cite{sm}.}
This parametrization form is inspired by the recent development of EEC 
in the nonperturbative collinear region~\cite{Lee:2025okn, Kang:2025zto, Chang:2025kgq, Guo:2025zwb, Kang:2026qbm}.
It reproduces the saturation behavior at small $\qf$
and power-law fall-off at large $\qf$~\cite{Zhu:2025qkx},
with a power of $2$ expected from perturbative matching
plus an ``anomalous'' dimension $\gamma_{\unpfec, f}(z)$.  
The transition in between occurs at the characteristic scale $\qf / z \sim \lambda_{\unpfec, f}$,
which is a constant for each flavor channel $f$.
The additional exponent $\kappa_{\unpfec,f}(z)$ fine-tunes the sharpness of this transition.
Treating $N_{\unpfec,f}(z)$, $\kappa_{\unpfec,f}(z)$, and $\gamma_{\unpfec,f}(z)$
as $z$-dependent interpolating functions
and fitting the FEC to their images extracted from {\tt Pythia} at $\mu\in \{ 5, 25, 50, 250, 500\}~\GeV$
produces good agreement,
as shown in \fig{fig:fec-dglap}(a),
with a clearly visible transition and plateau behavior at low $\qf$.

Similarly, we parametrize the Collins FEC at $\mu_0$ by
\begin{align}
	&\colfec_{f}(z, \qf^2; \mu_0)
	= N_{\colfec, f}(z) 
	\nn\\
	&\hspace{4em} \times 
	\frac{\qf}{z \lambda_{\colfec, f}}
	\biggbb{ \,
		1 + \biggpp{ \frac{\qf}{z \lambda_{\colfec, f} } }^{\kappa_{\colfec, f}}
	\, }^{-\frac{4 + \gamma_{\colfec, f}}{\kappa_{\colfec, f}}}.
\label{eq:fec-param-H}
\end{align}
Instead of reaching a plateau at small $\qf$, 
this is suppressed with the extra factor 
$\qf / (z \lambda_{\colfec, f})$. 
This is because azimuthal angle $\phi$ cannot be defined as $\theta \to 0$;
{this behavior is also expected from chiral perturbation theory.}
At large $\qf$, the Collins FEC matches onto twist-3 collinear fragmentation functions~\cite{Bacchetta:2008xw,Yuan:2009dw,Zhou:2011ba},
so scales as $\qf^{-3}$.
Unlike $\unpfec_f$, the Collins FEC $\colfec_{f}$ cannot be simulated 
using {\tt Pythia}, which does not incorporate the relevant physics. We therefore fix
$\gamma_{\colfec, f}$ and $\kappa_{\colfec, f}$ to the average values of the corresponding parameters in \eq{eq:fec-param-D}.
The normalization factor $N_{\colfec, f}(z)$ are qualitatively determined by an approximate relation to the Collins function $H_f^\perp(z, z^2 k_T^2)$~\cite{Liu:2024kqt,Cao:2025icu},
\begin{align}
	\int_0^{\infty} d\qf^2 \, \qf \, \colfec_{f}(z, \qf^2)
	&= - z^2 \int_0^{\infty} \frac{dk_T^2}{2m_h} k_T^2 \,
		H_f^\perp(z, z^2 k_T^2),
\label{eq:moment-relation}
\end{align}
point by point in $z$,
where $m_h$ is the mass of $h$
and the Collins function is taken from the {\tt JAM3D-22} fit~\cite{Gamberg:2022kdb}.
While \eq{eq:moment-relation} is established to apply at the bare distribution level and is subject to renormalization,
we regard it here solely as a reasonable constraint for modeling the FEC.
The $\qfcolcut$ remains as a free parameter, 
which we will scan to examine the effects of its relative size with $\lambda_{\unpfec}$.

As an example, we illustrate our models of Collins FECs in cumulative form in \fig{fig:fec-dglap}(b) to indicate their integrated strength,
with $\qfcolcut$ chosen at $3~\GeV$.
We also show their ratios to the unpolarized FEC,
which give the spin analyzing power of the fragmenting quark.

\emph{Universality.}---
The measurement of energy flow correlated inclusive hadron production 
can be easily extended to processes beyond $e^+e^-$ collisions.
In \refcite{Cao:2025icu}, we have provided a detailed study of this in $ep$ collisions,
$
	p^{\uparrow}(p) + e(\ell) \to e(\ell') + \bb{ h(p_h) + \E } + X,
$
with the proton beam being transversely polarized.
In the Breit frame, 
which is oriented such that the electron scatters in the $\hat x$-$\hat z$ plane
with $p$ moving in the $\hat z$ direction to collide head on with 
the virtual photon carrying momentum $q = \ell - \ell'$,
two types of observables are discussed~\cite{Cao:2025icu}, 
one with the transverse momentum of $h$ measured at a large value, and the other with it 
integrated over.

For the latter case, the $h$-$\E$ system is described by
$z = p \cdot p_h / p \cdot q$ of the hadron $h$
and $(\qf, \phi)$ of the nearby energy flow $\E$.
The hard scale of the process is provided by 
the longitudinal momentum of $h$, which is of order $z Q$,
with $Q = \sqrt{-q^2}$.
When $z Q \gg \qf$, the energy flow-weighted cross section $\Sigma$
can be factorized into parton distribution functions (PDFs) and FECs,
taking the form at LO~\cite{Cao:2025icu},
\begin{align}
	&\frac{d\Sigma_{ep\to ehX}}{dx_B \, dQ^2 \, d\phi_S \, dz \, d\qf^2 \, d\phi}
	\nn\\
	&\hspace{0.5em}
	= \frac{\alpha^2}{Q^4} \bb{ 1 + (1-y)^2 } \bigbb{ 
		U_{ep}
	 	+ S_T \, V_{ep} \sin(\phi-\phi_S)
	},
\label{eq:ep-xsec}
\end{align}
where $x_B = Q^2 / (2 p \cdot q)$ and,
with $y = p \cdot q / p \cdot \ell$,
\begin{align}
	U_{ep}
	&= \sum_{q = u, \bar u, \cdots} e_q^2 \,
		f_q(x_B) \, \unpfec_{h/q}(z, \qf^2),
	\nn\\
	V_{ep}
	&= \frac{2(1-y)}{ 1 + (1-y)^2 } 
        \sum_{q = u, \bar u, \cdots} e_q^2 \,
		h_q(x_B) \, \colfec_{h/q}(z, \qf^2),
\label{eq:ep-fact}
\end{align}
where $f_q$ and $h_q$ are the unpolarized and transversity PDFs, respectively.
The factorization scale $\mu$ is suppressed hereafter for notational simplicity.
The azimuthal correlation between the energy flow along $\phi$
and the proton transverse spin 
$\bm{S}_T = S_T \, (\cos\phi_S, \sin\phi_S)$ 
along $\phi_S$ is dynamically mediated 
by the transversity PDF and Collins FEC.

In this Letter, we further extend the FEC observable to $pp$ collisions 
with one of the beams transversely polarized,
$
	p^{\uparrow}(p_1) + p(p_2) \to \bb{ h(p_h) + \E } + X,
$
where we take the polarized proton $p^{\uparrow}$ to move along the $z$ direction, with its spin $\bm{S}_T$ fixed
along the $x$ direction in the lab frame,
and inclusively observe an energy flow $\E$
alongside a hadron $h$ with a high transverse momentum $\bm{p}_{hT} = p_{hT} \, (\cos\phi_h, \sin\phi_h)$.
When $p_{hT} \gg \qf$, this energy flow-weighted cross section $\Sigma$ also factorizes,
and at LO reads
\begin{align}
	\frac{d\Sigma_{pp\to hX}}{dp_{hT}^2 \, dy_h \, d\phi_h \, d\qf^2 \, d\phi}
	&= \frac{\alpha_s^2}{2s^2} \bb{
		U_{pp}
	 	+ S_T \, V_{pp}\sin(\phi+\phi_h)
	 }.
\label{eq:pp-xsec}
\end{align}
Here, $s = (p_1 + p_2)^2$, $\alpha_s$ is the strong coupling, $y_h$ is the rapidity of $h$, and
\begin{align}
	U_{pp}
	&= \sum_{a,b,c} \int \frac{dx_1 dx_2}{x_1^2 \, x_2^2 \, \bar z} 
		f_a(x_1) f_b(x_2)
		\, C_{abc}(\hat y)
		\, \unpfec_{h/c}(\bar z, \qf^2),
	\nn\\
	V_{pp}
	&= \sum_{a,b,c} \int \frac{dx_1 dx_2}{x_1^2 \, x_2^2 \, \bar z} 
		h_a(x_1) f_b(x_2)
		\, T_{abc}(\hat y)
		\, \colfec_{h/c}(\bar z, \qf^2),
\label{eq:pp-fact}
\end{align}
where $C$ and $T$ are the unpolarized and spin-transfer hard coefficients of the partonic scattering $ab \to cX$, 
whose explicit forms are collected in the SM~\cite{sm}.
The $\hat y = y_h - \ln (x_1 / x_2) / 2$ is the hadron rapidity in the partonic c.m.\ frame.
The momentum fraction of $h$ in $c$ has been
fixed at LO to $\bar z = 2 p_{hT} \cosh( \hat y ) / \sqrt{x_1 x_2 s}$.
By chiral symmetry, the sum over $a$ and $c$ in $V_{pp}$ is restricted to the ``non-singlet'' channel, with $a$ and $c$ being quarks or antiquarks of the same flavor.
Their spin transfer causes the azimuthal correlation 
$\sin(\phi + \phi_h)$
between the energy flow along $\phi$ and the hadron production plane along $\phi_h$.

The all-order FEC factorization in $ep$ collisions has been established in \refcite{Cao:2025icu}.
Extending to $pp$ collisions is nontrivial 
but can be achieved following the factorization of fragmentation functions~\cite{Nayak:2005rt}.
In particular, soft gluons pinched in the Glauber region are separated 
from the high-$p_{hT}$ hadron $h$ and energy flow $\E$,
so after the FEC is factorized out, 
the sum-over-cuts method in the Drell-Yan factorization~\cite{Bodwin:1984hc, Collins:1985ue, Collins:1988ig, Collins:2011zzd} 
can be applied in the same way to cancel all the soft poles in the Glauber region,
which allows the two beams to be factorized and all soft gluons canceled.
The important practical result is that the FECs factorized from $e^+e^-$, $ep$, and $pp$ collisions are the same objects 
given in \eqs{eq:fec-matrix}{eq:fec-spin}.

We have carried out a universality check for the unpolarized FEC using LO calculations and {\tt Pythia} simulations.
Specifically, we compare the {\tt Pythia} results for the unpolarized contributions in \eqs{eq:ep-xsec}{eq:pp-xsec} with the factorization predictions from \eqs{eq:ep-fact}{eq:pp-fact}, 
using the FECs extracted from $e^+e^-$ simulations and parametrized in \eq{eq:fec-param-D},
and have found overall good agreement~\cite{sm}.
While this comparison is not to replace a test with experimental data,
it does constitute a nontrivial check of both the universality of FECs
and our parametrization form.

\emph{Collins FEC asymmetry.}---
The Collins
effect of TMD observables has been measured in $e^+e^-$~\cite{Belle:2005dmx,Belle:2008fdv,BaBar:2013jdt}, 
$ep$~\cite{HERMES:2004mhh,HERMES:2010mmo,COMPASS:2012ozz},
and $pp$~\cite{STAR:2022hqg,STAR:2025xyp} collisions.
As indicated by Eqs.~\eqref{eq:ep-xsec}--\eqref{eq:pp-fact}, FECs provide a new type of Collins effect in the framework of collinear factorization, to which we refer as Collins FEC asymmetry. The observation does not require a back-to-back configuration or jet reconstruction, so is in principle simpler. In the following, we
present predictions for Collins FEC asymmetries
in all the three classes of collision processes,
based on the LO factorization formulas and our FEC modeling.

As shown in \fig{fig:fec-dglap}(c),
the perturbative to nonperturbative transition in the $\qf$ dependence of the Collins FEC 
is characterized by a hump around $\qf = z \qfcolcut$,
in contrast to the plateau behavior of $\unpfec$.
This feature is also observed in the recent RHIC measurement of the transverse spin asymmetry associated with energy-weighted hadrons in jets~\cite{STARCollaboration:2026kaa} 
and now emerges naturally from our FEC parametrization.
Using FECs eliminates systematic uncertainties associated with jets and facilitates the study of probing chiral symmetry breaking.

\begin{figure}[htbp]
  \centering
  \includegraphics[width=\linewidth]{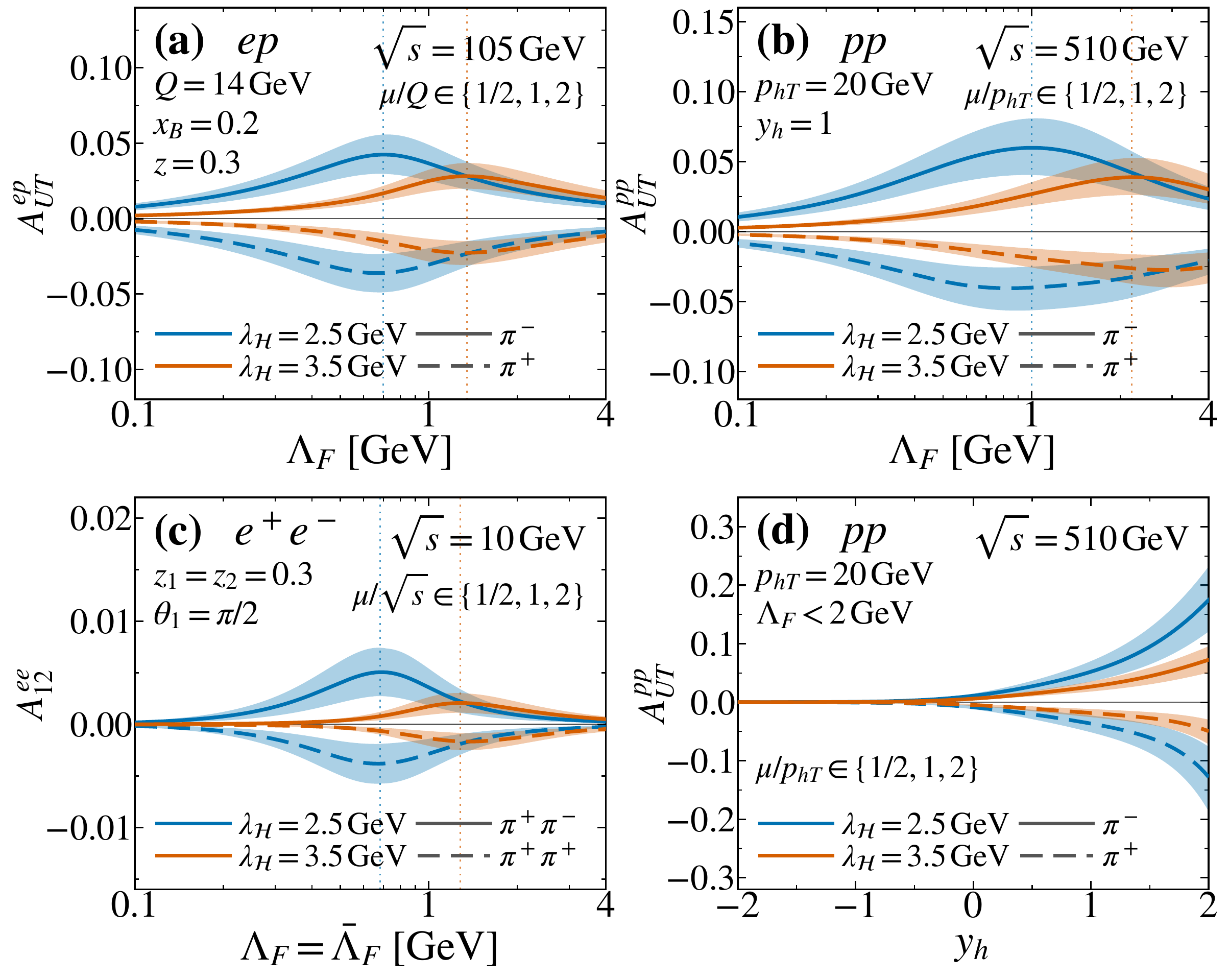}
  \caption{
  Collins FEC asymmetries in $ep$, $pp$, and $e^+e^-$ collisions, with two choices of $\qfcolcut$.
  (a)--(c) show their $\qf$ spectra at typical kinematic points of EIC, RHIC, and Belle, respectively. 
  (d) displays the $pp$ Collins FEC asymmetry as a function of the hadron rapidity $y_h$. 
  Shaded bands include uncertainties from FEC modeling and scale variations.
  }
  \label{fig:collins-asy}
\end{figure}

For both $ep$ and $pp$ collisions, the Collins FEC is probed via 
single transverse spin asymmetries, 
\begin{equation}
A_{UT}^i = V_i / U_i,
\qquad
i = ep, \, pp,
\end{equation}
given by \eqs{eq:ep-fact}{eq:pp-fact}, respectively.
These are shown in \fig{fig:collins-asy}(a) and (b)
for charged pion productions 
at typical kinematic points of EIC and RHIC.
The Collins transition scale $\lambda_{\colfec}$ is varied 
below and above the $\lambda_{\unpfec} \sim 2.9~\GeV$,
which is manifested by the hump position in the $\qf$ spectrum.
Since the azimuthally symmetric energy flow spectrum can be separately measured
to probe the confinement transition scale $\lambda_{\unpfec}$,
the azimuthal asymmetry 
provides an additional handle to scrutinize 
the chiral symmetry breaking scale $\lambda_{\colfec}$.

The Collins FEC asymmetry can also be studied at $e^+e^-$ collisions
via the azimuthal correlation of energy flows surrounding a pair of back-to-back hadrons. 
This does not rely on polarized beams 
and generalizes the single-hadron case in \eq{eq:ee-1pI-fec-lo}
to give
\begin{align}
	&\frac{d\Sigma_{ee \to h_1 h_2 X}}{
		dz_1 \, dz_2 \, d\cos\theta_1 \, d\qf^2 \, d\phi_1 \, d\bar{\Lambda}_F^2 \, d\phi_2}
	\nn\\
	&\hspace{2em}
	= \frac{3}{8} \pp{1 + \cos^2\theta_1 } \sigma_0 \bigbb{ U_{ee}
	 	+ V_{ee}\cos(\phi_2-\phi_1)
	},
\label{eq:ee-xsec}
\end{align}
where $\sigma_0$ is the same as in \eq{eq:ee-1pI-fec-lo}.
The $z_i = 2E_i / \sqrt{s}$ is defined for each of the two hadrons $h_i$ ($i = 1, 2$),
and $(\qf^2, \phi_1)$ and $(\bar{\Lambda}_F^2, \phi_2)$
describe their associated energy flows, respectively. 
$\theta_1$ is the polar angle of $h_1$ in the $e^+e^-$ c.m.\ frame.
The $U_{ee}$ and $V_{ee}$ can be again factorized into the same FECs,
and take the following form at LO,
\begin{align}
	U_{ee}
	&= \sum_{q = u, \bar u, \cdots} e_q^2\,
	\unpfec_{h_1/q}(z_1, \qf^2) \, \unpfec_{h_2/\bar q}(z_2, \bar{\Lambda}_F^2),
	\nn\\
	V_{ee}
	&= \frac{\sin^2\theta_1}{1 + \cos^2\theta_1 }
	\sum_{q = u, \bar u, \cdots} e_q^2\,
	\colfec_{h_1/q}(z_1, \qf^2) \, \colfec_{h_2/\bar q}(z_2, \bar{\Lambda}_F^2).
\label{eq:ee-fact}
\end{align}
The azimuthal correlation $\cos(\phi_2-\phi_1)$
resembles the Artru-Collins asymmetry for dihadron pair production~\cite{Artru:1995zu,Boer:2003ya},
which has been measured at Belle with appreciable precision~\cite{Belle:2011cur}.
The FEC counterpart, 
\begin{equation}
A_{12}^{ee} = V_{ee} / U_{ee},
\end{equation}
is shown in \fig{fig:collins-asy}(c) as a function of $\qf$,
with magnitude comparable with the measurement in \refcite{Belle:2011cur}.

It is evident from \eqs{eq:ep-fact}{eq:ee-fact} that 
the values of Collins FEC asymmetries, $A_{UT}^{ep}$ and $A_{12}^{ee}$,
are most significant at small $y$ and central angular regions, respectively.  
Not as obvious for the $pp$ case, \eq{eq:pp-fact}, 
the partonic scattering contains a $t$-channel enhancement 
for the asymmetry $A_{UT}^{pp}$,
so it can be best measured in the forward rapidity region,
as shown in \fig{fig:collins-asy}(d).
The magnitude is of the same order as the energy-flow-in-jets measurement in \refcite{STARCollaboration:2026kaa},
and hence we expect the Collins FEC asymmetry to be easily measurable as well.

\emph{Summary and prospects.}---
Our phenomenological study of energy-flow-accompanied inclusive hadron production highlights the potential of FECs as probes of the nonperturbative scales associated with confinement and, in particular, chiral symmetry breaking. 
Our predictions for the Collins asymmetries further indicate that the corresponding signals are within reach of current and future experimental facilities.
We have also thoroughly examined the universality property of the factorized FECs in $e^+e^-$, $ep$, and $pp$ collisions,
without the need for jet algorithms or simultaneous measurements of other heavy objects.
These features make FECs stand out among energy-flow-based observables and provide theoretically controlled handles on nonperturbative fragmentation. 
Furthermore, once the spin analyzing power of FECs of \fig{fig:fec-dglap} is precisely determined from experiments, it can be utilized to measure the quark spins in the final states of hard scattering processes. 
The nontrivial correlations of the final-state quark spins will provide a unique tool to unveil the quantum entanglement there~\cite{Cheng:2025cuv,Liu:2026dzv,Fucilla:2026zhy}.

\begin{acknowledgments}
We thank Ian Moult, Hao Chen, and Kyle Lee for helpful discussions.
This work is supported in part by the National Natural Science
Foundation of China under Grant Nos.~12235001 and 12425505,
the Fundamental and Interdisciplinary Disciplines Breakthrough Plan
of the Ministry of Education of China (No.~JYB2025XDXM204), and the Fundamental Research
Funds for the Central Universities at Peking University.
X.~Liu is supported by the National Natural Science Foundation of
China under Grant Nos.~12625505 and 12547109 and the Fundamental
Research Funds for the Central Universities at Beijing Normal University.
Z.~Yu is supported
by the U.S.\ Department of Energy under Contract No.\ DE-SC0012704.
C.-P.~Yuan is supported by the U.S. National Science Foundation
under Grant No.~PHY-2310291.
F.~Yuan is supported by the U.S. Department of Energy, Office of
Science, Office of Nuclear Physics, under Contract No.~DE-AC02-05CH11231.
\end{acknowledgments}

\FloatBarrier
\bibliographystyle{apsrev}
\bibliography{ref}

\begin{thebibliography}{101}
\expandafter\ifx\csname natexlab\endcsname\relax\def\natexlab#1{#1}\fi
\expandafter\ifx\csname bibnamefont\endcsname\relax
  \def\bibnamefont#1{#1}\fi
\expandafter\ifx\csname bibfnamefont\endcsname\relax
  \def\bibfnamefont#1{#1}\fi
\expandafter\ifx\csname citenamefont\endcsname\relax
  \def\citenamefont#1{#1}\fi
\expandafter\ifx\csname url\endcsname\relax
  \def\url#1{\texttt{#1}}\fi
\expandafter\ifx\csname urlprefix\endcsname\relax\def\urlprefix{URL }\fi
\providecommand{\bibinfo}[2]{#2}
\providecommand{\eprint}[2][]{\url{#2}}

\bibitem[{\citenamefont{Basham et~al.}(1978{\natexlab{a}})\citenamefont{Basham,
  Brown, Ellis, and Love}}]{Basham:1978bw}
\bibinfo{author}{\bibfnamefont{C.~L.} \bibnamefont{Basham}},
  \bibinfo{author}{\bibfnamefont{L.~S.} \bibnamefont{Brown}},
  \bibinfo{author}{\bibfnamefont{S.~D.} \bibnamefont{Ellis}}, \bibnamefont{and}
  \bibinfo{author}{\bibfnamefont{S.~T.} \bibnamefont{Love}},
  \bibinfo{journal}{Phys. Rev. Lett.} \textbf{\bibinfo{volume}{41}},
  \bibinfo{pages}{1585} (\bibinfo{year}{1978}{\natexlab{a}}).

\bibitem[{\citenamefont{Basham et~al.}(1979{\natexlab{a}})\citenamefont{Basham,
  Brown, Ellis, and Love}}]{Basham:1978zq}
\bibinfo{author}{\bibfnamefont{C.~L.} \bibnamefont{Basham}},
  \bibinfo{author}{\bibfnamefont{L.~S.} \bibnamefont{Brown}},
  \bibinfo{author}{\bibfnamefont{S.~D.} \bibnamefont{Ellis}}, \bibnamefont{and}
  \bibinfo{author}{\bibfnamefont{S.~T.} \bibnamefont{Love}},
  \bibinfo{journal}{Phys. Rev. D} \textbf{\bibinfo{volume}{19}},
  \bibinfo{pages}{2018} (\bibinfo{year}{1979}{\natexlab{a}}).

\bibitem[{\citenamefont{Basham et~al.}(1978{\natexlab{b}})\citenamefont{Basham,
  Brown, Ellis, and Love}}]{Basham:1977iq}
\bibinfo{author}{\bibfnamefont{C.~L.} \bibnamefont{Basham}},
  \bibinfo{author}{\bibfnamefont{L.~S.} \bibnamefont{Brown}},
  \bibinfo{author}{\bibfnamefont{S.~D.} \bibnamefont{Ellis}}, \bibnamefont{and}
  \bibinfo{author}{\bibfnamefont{S.~T.} \bibnamefont{Love}},
  \bibinfo{journal}{Phys. Rev. D} \textbf{\bibinfo{volume}{17}},
  \bibinfo{pages}{2298} (\bibinfo{year}{1978}{\natexlab{b}}).

\bibitem[{\citenamefont{Basham et~al.}(1979{\natexlab{b}})\citenamefont{Basham,
  Brown, Ellis, and Love}}]{Basham:1979gh}
\bibinfo{author}{\bibfnamefont{C.~L.} \bibnamefont{Basham}},
  \bibinfo{author}{\bibfnamefont{L.~S.} \bibnamefont{Brown}},
  \bibinfo{author}{\bibfnamefont{S.~D.} \bibnamefont{Ellis}}, \bibnamefont{and}
  \bibinfo{author}{\bibfnamefont{S.~T.} \bibnamefont{Love}},
  \bibinfo{journal}{Phys. Lett. B} \textbf{\bibinfo{volume}{85}},
  \bibinfo{pages}{297} (\bibinfo{year}{1979}{\natexlab{b}}).

\bibitem[{\citenamefont{Moult and Zhu}(2025)}]{Moult:2025nhu}
\bibinfo{author}{\bibfnamefont{I.}~\bibnamefont{Moult}} \bibnamefont{and}
  \bibinfo{author}{\bibfnamefont{H.~X.} \bibnamefont{Zhu}}
  (\bibinfo{year}{2025}), \eprint{2506.09119}.

\bibitem[{\citenamefont{Sterman}(1975)}]{Sterman:1975xv}
\bibinfo{author}{\bibfnamefont{G.~F.} \bibnamefont{Sterman}}
  (\bibinfo{year}{1975}).

\bibitem[{\citenamefont{Sveshnikov and Tkachov}(1996)}]{Sveshnikov:1995vi}
\bibinfo{author}{\bibfnamefont{N.~A.} \bibnamefont{Sveshnikov}}
  \bibnamefont{and} \bibinfo{author}{\bibfnamefont{F.~V.}
  \bibnamefont{Tkachov}}, \bibinfo{journal}{Phys. Lett. B}
  \textbf{\bibinfo{volume}{382}}, \bibinfo{pages}{403} (\bibinfo{year}{1996}),
  \eprint{hep-ph/9512370}.

\bibitem[{\citenamefont{Korchemsky and Sterman}(1999)}]{Korchemsky:1999kt}
\bibinfo{author}{\bibfnamefont{G.~P.} \bibnamefont{Korchemsky}}
  \bibnamefont{and} \bibinfo{author}{\bibfnamefont{G.~F.}
  \bibnamefont{Sterman}}, \bibinfo{journal}{Nucl. Phys. B}
  \textbf{\bibinfo{volume}{555}}, \bibinfo{pages}{335} (\bibinfo{year}{1999}),
  \eprint{hep-ph/9902341}.

\bibitem[{\citenamefont{Belitsky et~al.}(2001)\citenamefont{Belitsky,
  Korchemsky, and Sterman}}]{Belitsky:2001ij}
\bibinfo{author}{\bibfnamefont{A.~V.} \bibnamefont{Belitsky}},
  \bibinfo{author}{\bibfnamefont{G.~P.} \bibnamefont{Korchemsky}},
  \bibnamefont{and} \bibinfo{author}{\bibfnamefont{G.~F.}
  \bibnamefont{Sterman}}, \bibinfo{journal}{Phys. Lett. B}
  \textbf{\bibinfo{volume}{515}}, \bibinfo{pages}{297} (\bibinfo{year}{2001}),
  \eprint{hep-ph/0106308}.

\bibitem[{\citenamefont{Belitsky et~al.}(2014)\citenamefont{Belitsky,
  Hohenegger, Korchemsky, Sokatchev, and Zhiboedov}}]{Belitsky:2013xxa}
\bibinfo{author}{\bibfnamefont{A.~V.} \bibnamefont{Belitsky}},
  \bibinfo{author}{\bibfnamefont{S.}~\bibnamefont{Hohenegger}},
  \bibinfo{author}{\bibfnamefont{G.~P.} \bibnamefont{Korchemsky}},
  \bibinfo{author}{\bibfnamefont{E.}~\bibnamefont{Sokatchev}},
  \bibnamefont{and}
  \bibinfo{author}{\bibfnamefont{A.}~\bibnamefont{Zhiboedov}},
  \bibinfo{journal}{Nucl. Phys. B} \textbf{\bibinfo{volume}{884}},
  \bibinfo{pages}{305} (\bibinfo{year}{2014}), \eprint{1309.0769}.

\bibitem[{\citenamefont{Chang et~al.}(2013{\natexlab{a}})\citenamefont{Chang,
  Procura, Thaler, and Waalewijn}}]{Chang:2013rca}
\bibinfo{author}{\bibfnamefont{H.-M.} \bibnamefont{Chang}},
  \bibinfo{author}{\bibfnamefont{M.}~\bibnamefont{Procura}},
  \bibinfo{author}{\bibfnamefont{J.}~\bibnamefont{Thaler}}, \bibnamefont{and}
  \bibinfo{author}{\bibfnamefont{W.~J.} \bibnamefont{Waalewijn}},
  \bibinfo{journal}{Phys. Rev. Lett.} \textbf{\bibinfo{volume}{111}},
  \bibinfo{pages}{102002} (\bibinfo{year}{2013}{\natexlab{a}}),
  \eprint{1303.6637}.

\bibitem[{\citenamefont{Chang et~al.}(2013{\natexlab{b}})\citenamefont{Chang,
  Procura, Thaler, and Waalewijn}}]{Chang:2013iba}
\bibinfo{author}{\bibfnamefont{H.-M.} \bibnamefont{Chang}},
  \bibinfo{author}{\bibfnamefont{M.}~\bibnamefont{Procura}},
  \bibinfo{author}{\bibfnamefont{J.}~\bibnamefont{Thaler}}, \bibnamefont{and}
  \bibinfo{author}{\bibfnamefont{W.~J.} \bibnamefont{Waalewijn}},
  \bibinfo{journal}{Phys. Rev. D} \textbf{\bibinfo{volume}{88}},
  \bibinfo{pages}{034030} (\bibinfo{year}{2013}{\natexlab{b}}),
  \eprint{1306.6630}.

\bibitem[{\citenamefont{Jaarsma et~al.}(2023)\citenamefont{Jaarsma, Li, Moult,
  Waalewijn, and Zhu}}]{Jaarsma:2023ell}
\bibinfo{author}{\bibfnamefont{M.}~\bibnamefont{Jaarsma}},
  \bibinfo{author}{\bibfnamefont{Y.}~\bibnamefont{Li}},
  \bibinfo{author}{\bibfnamefont{I.}~\bibnamefont{Moult}},
  \bibinfo{author}{\bibfnamefont{W.~J.} \bibnamefont{Waalewijn}},
  \bibnamefont{and} \bibinfo{author}{\bibfnamefont{H.~X.} \bibnamefont{Zhu}},
  \bibinfo{journal}{JHEP} \textbf{\bibinfo{volume}{12}}, \bibinfo{pages}{087}
  (\bibinfo{year}{2023}), \eprint{2307.15739}.

\bibitem[{\citenamefont{Chen et~al.}(2023)\citenamefont{Chen, Jaarsma, Li,
  Moult, Waalewijn, and Zhu}}]{Chen:2022pdu}
\bibinfo{author}{\bibfnamefont{H.}~\bibnamefont{Chen}},
  \bibinfo{author}{\bibfnamefont{M.}~\bibnamefont{Jaarsma}},
  \bibinfo{author}{\bibfnamefont{Y.}~\bibnamefont{Li}},
  \bibinfo{author}{\bibfnamefont{I.}~\bibnamefont{Moult}},
  \bibinfo{author}{\bibfnamefont{W.~J.} \bibnamefont{Waalewijn}},
  \bibnamefont{and} \bibinfo{author}{\bibfnamefont{H.~X.} \bibnamefont{Zhu}},
  \bibinfo{journal}{JHEP} \textbf{\bibinfo{volume}{07}}, \bibinfo{pages}{185}
  (\bibinfo{year}{2023}), \eprint{2210.10058}.

\bibitem[{\citenamefont{Chen et~al.}(2025)\citenamefont{Chen, Jaarsma, Li,
  Moult, Waalewijn, and Zhu}}]{Chen:2022muj}
\bibinfo{author}{\bibfnamefont{H.}~\bibnamefont{Chen}},
  \bibinfo{author}{\bibfnamefont{M.}~\bibnamefont{Jaarsma}},
  \bibinfo{author}{\bibfnamefont{Y.}~\bibnamefont{Li}},
  \bibinfo{author}{\bibfnamefont{I.}~\bibnamefont{Moult}},
  \bibinfo{author}{\bibfnamefont{W.~J.} \bibnamefont{Waalewijn}},
  \bibnamefont{and} \bibinfo{author}{\bibfnamefont{H.~X.} \bibnamefont{Zhu}},
  \bibinfo{journal}{Phys. Rev. D} \textbf{\bibinfo{volume}{111}},
  \bibinfo{pages}{076021} (\bibinfo{year}{2025}), \eprint{2210.10061}.

\bibitem[{\citenamefont{Li et~al.}(2022)\citenamefont{Li, Moult, van Velzen,
  Waalewijn, and Zhu}}]{Li:2021zcf}
\bibinfo{author}{\bibfnamefont{Y.}~\bibnamefont{Li}},
  \bibinfo{author}{\bibfnamefont{I.}~\bibnamefont{Moult}},
  \bibinfo{author}{\bibfnamefont{S.~S.} \bibnamefont{van Velzen}},
  \bibinfo{author}{\bibfnamefont{W.~J.} \bibnamefont{Waalewijn}},
  \bibnamefont{and} \bibinfo{author}{\bibfnamefont{H.~X.} \bibnamefont{Zhu}},
  \bibinfo{journal}{Phys. Rev. Lett.} \textbf{\bibinfo{volume}{128}},
  \bibinfo{pages}{182001} (\bibinfo{year}{2022}), \eprint{2108.01674}.

\bibitem[{\citenamefont{Lee et~al.}(2024)\citenamefont{Lee, Moult, Ringer, and
  Waalewijn}}]{Lee:2023xzv}
\bibinfo{author}{\bibfnamefont{K.}~\bibnamefont{Lee}},
  \bibinfo{author}{\bibfnamefont{I.}~\bibnamefont{Moult}},
  \bibinfo{author}{\bibfnamefont{F.}~\bibnamefont{Ringer}}, \bibnamefont{and}
  \bibinfo{author}{\bibfnamefont{W.~J.} \bibnamefont{Waalewijn}},
  \bibinfo{journal}{JHEP} \textbf{\bibinfo{volume}{01}}, \bibinfo{pages}{194}
  (\bibinfo{year}{2024}), \eprint{2308.00028}.

\bibitem[{\citenamefont{Lee et~al.}(2026{\natexlab{a}})\citenamefont{Lee,
  Moult, and Waalewijn}}]{Lee:2026hub}
\bibinfo{author}{\bibfnamefont{K.}~\bibnamefont{Lee}},
  \bibinfo{author}{\bibfnamefont{I.}~\bibnamefont{Moult}}, \bibnamefont{and}
  \bibinfo{author}{\bibfnamefont{W.~J.} \bibnamefont{Waalewijn}}
  (\bibinfo{year}{2026}{\natexlab{a}}), \eprint{2607.00087}.

\bibitem[{\citenamefont{Hofman and Maldacena}(2008)}]{Hofman:2008ar}
\bibinfo{author}{\bibfnamefont{D.~M.} \bibnamefont{Hofman}} \bibnamefont{and}
  \bibinfo{author}{\bibfnamefont{J.}~\bibnamefont{Maldacena}},
  \bibinfo{journal}{JHEP} \textbf{\bibinfo{volume}{05}}, \bibinfo{pages}{012}
  (\bibinfo{year}{2008}), \eprint{0803.1467}.

\bibitem[{\citenamefont{Kravchuk and Simmons-Duffin}(2018)}]{Kravchuk:2018htv}
\bibinfo{author}{\bibfnamefont{P.}~\bibnamefont{Kravchuk}} \bibnamefont{and}
  \bibinfo{author}{\bibfnamefont{D.}~\bibnamefont{Simmons-Duffin}},
  \bibinfo{journal}{JHEP} \textbf{\bibinfo{volume}{11}}, \bibinfo{pages}{102}
  (\bibinfo{year}{2018}), \eprint{1805.00098}.

\bibitem[{\citenamefont{Kologlu et~al.}(2021)\citenamefont{Kologlu, Kravchuk,
  Simmons-Duffin, and Zhiboedov}}]{Kologlu:2019mfz}
\bibinfo{author}{\bibfnamefont{M.}~\bibnamefont{Kologlu}},
  \bibinfo{author}{\bibfnamefont{P.}~\bibnamefont{Kravchuk}},
  \bibinfo{author}{\bibfnamefont{D.}~\bibnamefont{Simmons-Duffin}},
  \bibnamefont{and}
  \bibinfo{author}{\bibfnamefont{A.}~\bibnamefont{Zhiboedov}},
  \bibinfo{journal}{JHEP} \textbf{\bibinfo{volume}{01}}, \bibinfo{pages}{128}
  (\bibinfo{year}{2021}), \eprint{1905.01311}.

\bibitem[{\citenamefont{Chang et~al.}(2022)\citenamefont{Chang, Kologlu,
  Kravchuk, Simmons-Duffin, and Zhiboedov}}]{Chang:2020qpj}
\bibinfo{author}{\bibfnamefont{C.-H.} \bibnamefont{Chang}},
  \bibinfo{author}{\bibfnamefont{M.}~\bibnamefont{Kologlu}},
  \bibinfo{author}{\bibfnamefont{P.}~\bibnamefont{Kravchuk}},
  \bibinfo{author}{\bibfnamefont{D.}~\bibnamefont{Simmons-Duffin}},
  \bibnamefont{and}
  \bibinfo{author}{\bibfnamefont{A.}~\bibnamefont{Zhiboedov}},
  \bibinfo{journal}{JHEP} \textbf{\bibinfo{volume}{05}}, \bibinfo{pages}{059}
  (\bibinfo{year}{2022}), \eprint{2010.04726}.

\bibitem[{\citenamefont{Chen et~al.}(2022{\natexlab{a}})\citenamefont{Chen,
  Moult, Sandor, and Zhu}}]{Chen:2022jhb}
\bibinfo{author}{\bibfnamefont{H.}~\bibnamefont{Chen}},
  \bibinfo{author}{\bibfnamefont{I.}~\bibnamefont{Moult}},
  \bibinfo{author}{\bibfnamefont{J.}~\bibnamefont{Sandor}}, \bibnamefont{and}
  \bibinfo{author}{\bibfnamefont{H.~X.} \bibnamefont{Zhu}},
  \bibinfo{journal}{JHEP} \textbf{\bibinfo{volume}{09}}, \bibinfo{pages}{199}
  (\bibinfo{year}{2022}{\natexlab{a}}), \eprint{2202.04085}.

\bibitem[{\citenamefont{Caron-Huot et~al.}(2023)\citenamefont{Caron-Huot,
  Kologlu, Kravchuk, Meltzer, and Simmons-Duffin}}]{Caron-Huot:2022eqs}
\bibinfo{author}{\bibfnamefont{S.}~\bibnamefont{Caron-Huot}},
  \bibinfo{author}{\bibfnamefont{M.}~\bibnamefont{Kologlu}},
  \bibinfo{author}{\bibfnamefont{P.}~\bibnamefont{Kravchuk}},
  \bibinfo{author}{\bibfnamefont{D.}~\bibnamefont{Meltzer}}, \bibnamefont{and}
  \bibinfo{author}{\bibfnamefont{D.}~\bibnamefont{Simmons-Duffin}},
  \bibinfo{journal}{JHEP} \textbf{\bibinfo{volume}{04}}, \bibinfo{pages}{014}
  (\bibinfo{year}{2023}), \eprint{2209.00008}.

\bibitem[{\citenamefont{Chen et~al.}(2021)\citenamefont{Chen, Moult, and
  Zhu}}]{Chen:2020adz}
\bibinfo{author}{\bibfnamefont{H.}~\bibnamefont{Chen}},
  \bibinfo{author}{\bibfnamefont{I.}~\bibnamefont{Moult}}, \bibnamefont{and}
  \bibinfo{author}{\bibfnamefont{H.~X.} \bibnamefont{Zhu}},
  \bibinfo{journal}{Phys. Rev. Lett.} \textbf{\bibinfo{volume}{126}},
  \bibinfo{pages}{112003} (\bibinfo{year}{2021}), \eprint{2011.02492}.

\bibitem[{\citenamefont{Chen et~al.}(2022{\natexlab{b}})\citenamefont{Chen,
  Moult, and Zhu}}]{Chen:2021gdk}
\bibinfo{author}{\bibfnamefont{H.}~\bibnamefont{Chen}},
  \bibinfo{author}{\bibfnamefont{I.}~\bibnamefont{Moult}}, \bibnamefont{and}
  \bibinfo{author}{\bibfnamefont{H.~X.} \bibnamefont{Zhu}},
  \bibinfo{journal}{JHEP} \textbf{\bibinfo{volume}{08}}, \bibinfo{pages}{233}
  (\bibinfo{year}{2022}{\natexlab{b}}), \eprint{2104.00009}.

\bibitem[{\citenamefont{Chen}(2024)}]{Chen:2023zzh}
\bibinfo{author}{\bibfnamefont{H.}~\bibnamefont{Chen}}, \bibinfo{journal}{JHEP}
  \textbf{\bibinfo{volume}{01}}, \bibinfo{pages}{035} (\bibinfo{year}{2024}),
  \eprint{2311.00350}.

\bibitem[{\citenamefont{Dixon et~al.}(2019)\citenamefont{Dixon, Moult, and
  Zhu}}]{Dixon:2019uzg}
\bibinfo{author}{\bibfnamefont{L.~J.} \bibnamefont{Dixon}},
  \bibinfo{author}{\bibfnamefont{I.}~\bibnamefont{Moult}}, \bibnamefont{and}
  \bibinfo{author}{\bibfnamefont{H.~X.} \bibnamefont{Zhu}},
  \bibinfo{journal}{Phys. Rev. D} \textbf{\bibinfo{volume}{100}},
  \bibinfo{pages}{014009} (\bibinfo{year}{2019}), \eprint{1905.01310}.

\bibitem[{\citenamefont{Chen et~al.}(2020)\citenamefont{Chen, Moult, Zhang, and
  Zhu}}]{Chen:2020vvp}
\bibinfo{author}{\bibfnamefont{H.}~\bibnamefont{Chen}},
  \bibinfo{author}{\bibfnamefont{I.}~\bibnamefont{Moult}},
  \bibinfo{author}{\bibfnamefont{X.}~\bibnamefont{Zhang}}, \bibnamefont{and}
  \bibinfo{author}{\bibfnamefont{H.~X.} \bibnamefont{Zhu}},
  \bibinfo{journal}{Phys. Rev. D} \textbf{\bibinfo{volume}{102}},
  \bibinfo{pages}{054012} (\bibinfo{year}{2020}), \eprint{2004.11381}.

\bibitem[{\citenamefont{Lee et~al.}(2025)\citenamefont{Lee, Me{\c{c}}aj, and
  Moult}}]{Lee:2022uwt}
\bibinfo{author}{\bibfnamefont{K.}~\bibnamefont{Lee}},
  \bibinfo{author}{\bibfnamefont{B.}~\bibnamefont{Me{\c{c}}aj}},
  \bibnamefont{and} \bibinfo{author}{\bibfnamefont{I.}~\bibnamefont{Moult}},
  \bibinfo{journal}{Phys. Rev. D} \textbf{\bibinfo{volume}{111}},
  \bibinfo{pages}{L011502} (\bibinfo{year}{2025}), \eprint{2205.03414}.

\bibitem[{\citenamefont{Chen et~al.}(2024)\citenamefont{Chen, Gao, Li, Xu,
  Zhang, and Zhu}}]{Chen:2023zlx}
\bibinfo{author}{\bibfnamefont{W.}~\bibnamefont{Chen}},
  \bibinfo{author}{\bibfnamefont{J.}~\bibnamefont{Gao}},
  \bibinfo{author}{\bibfnamefont{Y.}~\bibnamefont{Li}},
  \bibinfo{author}{\bibfnamefont{Z.}~\bibnamefont{Xu}},
  \bibinfo{author}{\bibfnamefont{X.}~\bibnamefont{Zhang}}, \bibnamefont{and}
  \bibinfo{author}{\bibfnamefont{H.~X.} \bibnamefont{Zhu}},
  \bibinfo{journal}{JHEP} \textbf{\bibinfo{volume}{05}}, \bibinfo{pages}{043}
  (\bibinfo{year}{2024}), \eprint{2307.07510}.

\bibitem[{\citenamefont{Lee et~al.}(2026{\natexlab{b}})\citenamefont{Lee, Li,
  Xu, and Zhang}}]{Lee:2026zyl}
\bibinfo{author}{\bibfnamefont{K.}~\bibnamefont{Lee}},
  \bibinfo{author}{\bibfnamefont{Y.}~\bibnamefont{Li}},
  \bibinfo{author}{\bibfnamefont{Z.}~\bibnamefont{Xu}}, \bibnamefont{and}
  \bibinfo{author}{\bibfnamefont{X.}~\bibnamefont{Zhang}}
  (\bibinfo{year}{2026}{\natexlab{b}}), \eprint{2606.02714}.

\bibitem[{\citenamefont{Liu et~al.}(2025)\citenamefont{Liu, Vogelsang, Yuan,
  and Zhu}}]{Liu:2024lxy}
\bibinfo{author}{\bibfnamefont{X.}~\bibnamefont{Liu}},
  \bibinfo{author}{\bibfnamefont{W.}~\bibnamefont{Vogelsang}},
  \bibinfo{author}{\bibfnamefont{F.}~\bibnamefont{Yuan}}, \bibnamefont{and}
  \bibinfo{author}{\bibfnamefont{H.~X.} \bibnamefont{Zhu}},
  \bibinfo{journal}{Phys. Rev. Lett.} \textbf{\bibinfo{volume}{134}},
  \bibinfo{pages}{151901} (\bibinfo{year}{2025}), \eprint{2410.16371}.

\bibitem[{\citenamefont{Barata et~al.}(2025)\citenamefont{Barata, Kang,
  Mayo~L{\'o}pez, and Penttala}}]{Barata:2024wsu}
\bibinfo{author}{\bibfnamefont{J.}~\bibnamefont{Barata}},
  \bibinfo{author}{\bibfnamefont{Z.-B.} \bibnamefont{Kang}},
  \bibinfo{author}{\bibfnamefont{X.}~\bibnamefont{Mayo~L{\'o}pez}},
  \bibnamefont{and} \bibinfo{author}{\bibfnamefont{J.}~\bibnamefont{Penttala}},
  \bibinfo{journal}{Phys. Rev. Lett.} \textbf{\bibinfo{volume}{134}},
  \bibinfo{pages}{251903} (\bibinfo{year}{2025}), \eprint{2411.11782}.

\bibitem[{\citenamefont{Lee and Stewart}(2026)}]{Lee:2025okn}
\bibinfo{author}{\bibfnamefont{K.}~\bibnamefont{Lee}} \bibnamefont{and}
  \bibinfo{author}{\bibfnamefont{I.~W.} \bibnamefont{Stewart}},
  \bibinfo{journal}{Phys. Rev. Lett.} \textbf{\bibinfo{volume}{136}},
  \bibinfo{pages}{081902} (\bibinfo{year}{2026}), \eprint{2507.11495}.

\bibitem[{\citenamefont{Guo et~al.}(2026{\natexlab{a}})\citenamefont{Guo, Yuan,
  and Zhao}}]{Guo:2025zwb}
\bibinfo{author}{\bibfnamefont{Y.}~\bibnamefont{Guo}},
  \bibinfo{author}{\bibfnamefont{F.}~\bibnamefont{Yuan}}, \bibnamefont{and}
  \bibinfo{author}{\bibfnamefont{W.}~\bibnamefont{Zhao}},
  \bibinfo{journal}{Phys. Rev. Lett.} \textbf{\bibinfo{volume}{136}},
  \bibinfo{pages}{081904} (\bibinfo{year}{2026}{\natexlab{a}}),
  \eprint{2507.15820}.

\bibitem[{\citenamefont{Chang et~al.}(2026)\citenamefont{Chang, Chen, Liu,
  Simmons-Duffin, Yuan, and Zhu}}]{Chang:2025kgq}
\bibinfo{author}{\bibfnamefont{C.-H.} \bibnamefont{Chang}},
  \bibinfo{author}{\bibfnamefont{H.}~\bibnamefont{Chen}},
  \bibinfo{author}{\bibfnamefont{X.}~\bibnamefont{Liu}},
  \bibinfo{author}{\bibfnamefont{D.}~\bibnamefont{Simmons-Duffin}},
  \bibinfo{author}{\bibfnamefont{F.}~\bibnamefont{Yuan}}, \bibnamefont{and}
  \bibinfo{author}{\bibfnamefont{H.~X.} \bibnamefont{Zhu}},
  \bibinfo{journal}{Phys. Rev. Lett.} \textbf{\bibinfo{volume}{136}},
  \bibinfo{pages}{081903} (\bibinfo{year}{2026}), \eprint{2507.15923}.

\bibitem[{\citenamefont{Kang et~al.}(2026{\natexlab{a}})\citenamefont{Kang,
  Metz, Pitonyak, and Zhang}}]{Kang:2025zto}
\bibinfo{author}{\bibfnamefont{Z.-B.} \bibnamefont{Kang}},
  \bibinfo{author}{\bibfnamefont{A.}~\bibnamefont{Metz}},
  \bibinfo{author}{\bibfnamefont{D.}~\bibnamefont{Pitonyak}}, \bibnamefont{and}
  \bibinfo{author}{\bibfnamefont{C.}~\bibnamefont{Zhang}},
  \bibinfo{journal}{Phys. Rev. Lett.} \textbf{\bibinfo{volume}{136}},
  \bibinfo{pages}{081905} (\bibinfo{year}{2026}{\natexlab{a}}),
  \eprint{2507.17444}.

\bibitem[{\citenamefont{Herrmann et~al.}(2025)\citenamefont{Herrmann, Kang,
  Penttala, and Zhang}}]{Herrmann:2025fqy}
\bibinfo{author}{\bibfnamefont{E.}~\bibnamefont{Herrmann}},
  \bibinfo{author}{\bibfnamefont{Z.-B.} \bibnamefont{Kang}},
  \bibinfo{author}{\bibfnamefont{J.}~\bibnamefont{Penttala}}, \bibnamefont{and}
  \bibinfo{author}{\bibfnamefont{C.}~\bibnamefont{Zhang}}
  (\bibinfo{year}{2025}), \eprint{2507.17704}.

\bibitem[{\citenamefont{Guo and Yuan}(2026)}]{Guo:2026bhd}
\bibinfo{author}{\bibfnamefont{Y.}~\bibnamefont{Guo}} \bibnamefont{and}
  \bibinfo{author}{\bibfnamefont{F.}~\bibnamefont{Yuan}}
  (\bibinfo{year}{2026}), \eprint{2607.09860}.

\bibitem[{\citenamefont{Komiske et~al.}(2023)\citenamefont{Komiske, Moult,
  Thaler, and Zhu}}]{Komiske:2022enw}
\bibinfo{author}{\bibfnamefont{P.~T.} \bibnamefont{Komiske}},
  \bibinfo{author}{\bibfnamefont{I.}~\bibnamefont{Moult}},
  \bibinfo{author}{\bibfnamefont{J.}~\bibnamefont{Thaler}}, \bibnamefont{and}
  \bibinfo{author}{\bibfnamefont{H.~X.} \bibnamefont{Zhu}},
  \bibinfo{journal}{Phys. Rev. Lett.} \textbf{\bibinfo{volume}{130}},
  \bibinfo{pages}{051901} (\bibinfo{year}{2023}), \eprint{2201.07800}.

\bibitem[{\citenamefont{Hayrapetyan et~al.}(2024)}]{CMS:2024mlf}
\bibinfo{author}{\bibfnamefont{A.}~\bibnamefont{Hayrapetyan}}
  \bibnamefont{et~al.} (\bibinfo{collaboration}{CMS}), \bibinfo{journal}{Phys.
  Rev. Lett.} \textbf{\bibinfo{volume}{133}}, \bibinfo{pages}{071903}
  (\bibinfo{year}{2024}), \eprint{2402.13864}.

\bibitem[{CMS(2025)}]{CMS-PAS-HIN-24-019}
\bibinfo{type}{Tech. Rep.}, \bibinfo{institution}{CERN},
  \bibinfo{address}{Geneva} (\bibinfo{year}{2025}),
  \urlprefix\url{https://cds.cern.ch/record/2929460}.

\bibitem[{\citenamefont{Bossi
  et~al.}(2025)}]{Electron-PositronAlliance:2025fhk}
\bibinfo{author}{\bibfnamefont{H.}~\bibnamefont{Bossi}} \bibnamefont{et~al.}
  (\bibinfo{collaboration}{Electron-Positron Alliance}) (\bibinfo{year}{2025}),
  \eprint{2511.00149}.

\bibitem[{\citenamefont{Acharya et~al.}(2026)}]{ALICE:2024dfl}
\bibinfo{author}{\bibfnamefont{S.}~\bibnamefont{Acharya}} \bibnamefont{et~al.}
  (\bibinfo{collaboration}{ALICE}), \bibinfo{journal}{Phys. Rev. D}
  \textbf{\bibinfo{volume}{114}}, \bibinfo{pages}{032002}
  (\bibinfo{year}{2026}), \eprint{2409.12687}.

\bibitem[{\citenamefont{Aboona et~al.}(2025{\natexlab{a}})}]{STAR:2025jut}
\bibinfo{author}{\bibfnamefont{B.~E.} \bibnamefont{Aboona}}
  \bibnamefont{et~al.} (\bibinfo{collaboration}{STAR}), \bibinfo{journal}{Phys.
  Rev. Lett.} \textbf{\bibinfo{volume}{135}}, \bibinfo{pages}{111901}
  (\bibinfo{year}{2025}{\natexlab{a}}), \eprint{2502.15925}.

\bibitem[{\citenamefont{Guo et~al.}(2026{\natexlab{b}})\citenamefont{Guo,
  Vogelsang, Yuan, and Zhao}}]{Guo:2025qnz}
\bibinfo{author}{\bibfnamefont{Y.}~\bibnamefont{Guo}},
  \bibinfo{author}{\bibfnamefont{W.}~\bibnamefont{Vogelsang}},
  \bibinfo{author}{\bibfnamefont{F.}~\bibnamefont{Yuan}}, \bibnamefont{and}
  \bibinfo{author}{\bibfnamefont{W.}~\bibnamefont{Zhao}},
  \bibinfo{journal}{Phys. Rev. D} \textbf{\bibinfo{volume}{113}},
  \bibinfo{pages}{074027} (\bibinfo{year}{2026}{\natexlab{b}}),
  \eprint{2512.15896}.

\bibitem[{CMS(2026)}]{CMS-PAS-SMP-25-009}
\bibinfo{type}{Tech. Rep.}, \bibinfo{institution}{CERN},
  \bibinfo{address}{Geneva} (\bibinfo{year}{2026}),
  \urlprefix\url{https://cds.cern.ch/record/2968132}.

\bibitem[{\citenamefont{Generet et~al.}(2026)\citenamefont{Generet, Lee, Moult,
  Poncelet, and Zhang}}]{Generet:2026vto}
\bibinfo{author}{\bibfnamefont{T.}~\bibnamefont{Generet}},
  \bibinfo{author}{\bibfnamefont{K.}~\bibnamefont{Lee}},
  \bibinfo{author}{\bibfnamefont{I.}~\bibnamefont{Moult}},
  \bibinfo{author}{\bibfnamefont{R.}~\bibnamefont{Poncelet}}, \bibnamefont{and}
  \bibinfo{author}{\bibfnamefont{X.}~\bibnamefont{Zhang}}
  (\bibinfo{year}{2026}), \eprint{2608.14793}.

\bibitem[{\citenamefont{Liu and Zhu}(2024)}]{Liu:2024kqt}
\bibinfo{author}{\bibfnamefont{X.}~\bibnamefont{Liu}} \bibnamefont{and}
  \bibinfo{author}{\bibfnamefont{H.~X.} \bibnamefont{Zhu}}
  (\bibinfo{year}{2024}), \eprint{2403.08874}.

\bibitem[{\citenamefont{Cao et~al.}(2026)\citenamefont{Cao, Yu, Yuan, Zhang,
  and Zhu}}]{Cao:2025icu}
\bibinfo{author}{\bibfnamefont{Q.-H.} \bibnamefont{Cao}},
  \bibinfo{author}{\bibfnamefont{Z.}~\bibnamefont{Yu}},
  \bibinfo{author}{\bibfnamefont{C.~P.} \bibnamefont{Yuan}},
  \bibinfo{author}{\bibfnamefont{S.}~\bibnamefont{Zhang}}, \bibnamefont{and}
  \bibinfo{author}{\bibfnamefont{H.~X.} \bibnamefont{Zhu}},
  \bibinfo{journal}{JHEP} \textbf{\bibinfo{volume}{02}}, \bibinfo{pages}{244}
  (\bibinfo{year}{2026}), \eprint{2509.18892}.

\bibitem[{\citenamefont{Zhu}(2026)}]{Zhu:2025qkx}
\bibinfo{author}{\bibfnamefont{Y.~J.} \bibnamefont{Zhu}},
  \bibinfo{journal}{Phys. Rev. D} \textbf{\bibinfo{volume}{113}},
  \bibinfo{pages}{014025} (\bibinfo{year}{2026}), \eprint{2509.01652}.

\bibitem[{\citenamefont{Collins}(1993)}]{Collins:1992kk}
\bibinfo{author}{\bibfnamefont{J.~C.} \bibnamefont{Collins}},
  \bibinfo{journal}{Nucl. Phys. B} \textbf{\bibinfo{volume}{396}},
  \bibinfo{pages}{161} (\bibinfo{year}{1993}), \eprint{hep-ph/9208213}.

\bibitem[{\citenamefont{Bacchetta et~al.}(2002)\citenamefont{Bacchetta, Kundu,
  Metz, and Mulders}}]{Bacchetta:2002tk}
\bibinfo{author}{\bibfnamefont{A.}~\bibnamefont{Bacchetta}},
  \bibinfo{author}{\bibfnamefont{R.}~\bibnamefont{Kundu}},
  \bibinfo{author}{\bibfnamefont{A.}~\bibnamefont{Metz}}, \bibnamefont{and}
  \bibinfo{author}{\bibfnamefont{P.~J.} \bibnamefont{Mulders}},
  \bibinfo{journal}{Phys. Rev. D} \textbf{\bibinfo{volume}{65}},
  \bibinfo{pages}{094021} (\bibinfo{year}{2002}), \eprint{hep-ph/0201091}.

\bibitem[{\citenamefont{Amrath et~al.}(2005)\citenamefont{Amrath, Bacchetta,
  and Metz}}]{Amrath:2005gv}
\bibinfo{author}{\bibfnamefont{D.}~\bibnamefont{Amrath}},
  \bibinfo{author}{\bibfnamefont{A.}~\bibnamefont{Bacchetta}},
  \bibnamefont{and} \bibinfo{author}{\bibfnamefont{A.}~\bibnamefont{Metz}},
  \bibinfo{journal}{Phys. Rev. D} \textbf{\bibinfo{volume}{71}},
  \bibinfo{pages}{114018} (\bibinfo{year}{2005}), \eprint{hep-ph/0504124}.

\bibitem[{\citenamefont{Boer}(2001)}]{Boer:2001he}
\bibinfo{author}{\bibfnamefont{D.}~\bibnamefont{Boer}}, \bibinfo{journal}{Nucl.
  Phys. B} \textbf{\bibinfo{volume}{603}}, \bibinfo{pages}{195}
  (\bibinfo{year}{2001}), \eprint{hep-ph/0102071}.

\bibitem[{\citenamefont{Kang et~al.}(2016)\citenamefont{Kang, Prokudin, Sun,
  and Yuan}}]{Kang:2015msa}
\bibinfo{author}{\bibfnamefont{Z.-B.} \bibnamefont{Kang}},
  \bibinfo{author}{\bibfnamefont{A.}~\bibnamefont{Prokudin}},
  \bibinfo{author}{\bibfnamefont{P.}~\bibnamefont{Sun}}, \bibnamefont{and}
  \bibinfo{author}{\bibfnamefont{F.}~\bibnamefont{Yuan}},
  \bibinfo{journal}{Phys. Rev. D} \textbf{\bibinfo{volume}{93}},
  \bibinfo{pages}{014009} (\bibinfo{year}{2016}), \eprint{1505.05589}.

\bibitem[{\citenamefont{Hatta et~al.}(2021{\natexlab{a}})\citenamefont{Hatta,
  Xiao, Yuan, and Zhou}}]{Hatta:2020bgy}
\bibinfo{author}{\bibfnamefont{Y.}~\bibnamefont{Hatta}},
  \bibinfo{author}{\bibfnamefont{B.-W.} \bibnamefont{Xiao}},
  \bibinfo{author}{\bibfnamefont{F.}~\bibnamefont{Yuan}}, \bibnamefont{and}
  \bibinfo{author}{\bibfnamefont{J.}~\bibnamefont{Zhou}},
  \bibinfo{journal}{Phys. Rev. Lett.} \textbf{\bibinfo{volume}{126}},
  \bibinfo{pages}{142001} (\bibinfo{year}{2021}{\natexlab{a}}),
  \eprint{2010.10774}.

\bibitem[{\citenamefont{Hatta et~al.}(2021{\natexlab{b}})\citenamefont{Hatta,
  Xiao, Yuan, and Zhou}}]{Hatta:2021jcd}
\bibinfo{author}{\bibfnamefont{Y.}~\bibnamefont{Hatta}},
  \bibinfo{author}{\bibfnamefont{B.-W.} \bibnamefont{Xiao}},
  \bibinfo{author}{\bibfnamefont{F.}~\bibnamefont{Yuan}}, \bibnamefont{and}
  \bibinfo{author}{\bibfnamefont{J.}~\bibnamefont{Zhou}},
  \bibinfo{journal}{Phys. Rev. D} \textbf{\bibinfo{volume}{104}},
  \bibinfo{pages}{054037} (\bibinfo{year}{2021}{\natexlab{b}}),
  \eprint{2106.05307}.

\bibitem[{\citenamefont{Bacchetta et~al.}(2001)\citenamefont{Bacchetta, Kundu,
  Metz, and Mulders}}]{Bacchetta:2001di}
\bibinfo{author}{\bibfnamefont{A.}~\bibnamefont{Bacchetta}},
  \bibinfo{author}{\bibfnamefont{R.}~\bibnamefont{Kundu}},
  \bibinfo{author}{\bibfnamefont{A.}~\bibnamefont{Metz}}, \bibnamefont{and}
  \bibinfo{author}{\bibfnamefont{P.~J.} \bibnamefont{Mulders}},
  \bibinfo{journal}{Phys. Lett. B} \textbf{\bibinfo{volume}{506}},
  \bibinfo{pages}{155} (\bibinfo{year}{2001}), \eprint{hep-ph/0102278}.

\bibitem[{\citenamefont{Yuan}(2008{\natexlab{a}})}]{Yuan:2008yv}
\bibinfo{author}{\bibfnamefont{F.}~\bibnamefont{Yuan}}, \bibinfo{journal}{Phys.
  Rev. D} \textbf{\bibinfo{volume}{77}}, \bibinfo{pages}{074019}
  (\bibinfo{year}{2008}{\natexlab{a}}), \eprint{0801.3441}.

\bibitem[{\citenamefont{Li et~al.}(2023)\citenamefont{Li, Liu, Yuan, and
  Zhu}}]{Li:2023gkh}
\bibinfo{author}{\bibfnamefont{X.~L.} \bibnamefont{Li}},
  \bibinfo{author}{\bibfnamefont{X.}~\bibnamefont{Liu}},
  \bibinfo{author}{\bibfnamefont{F.}~\bibnamefont{Yuan}}, \bibnamefont{and}
  \bibinfo{author}{\bibfnamefont{H.~X.} \bibnamefont{Zhu}},
  \bibinfo{journal}{Phys. Rev. D} \textbf{\bibinfo{volume}{108}},
  \bibinfo{pages}{L091502} (\bibinfo{year}{2023}), \eprint{2308.10942}.

\bibitem[{\citenamefont{Guo et~al.}(2025{\natexlab{a}})\citenamefont{Guo, Liu,
  Yuan, and Zhu}}]{Guo:2024jch}
\bibinfo{author}{\bibfnamefont{Y.}~\bibnamefont{Guo}},
  \bibinfo{author}{\bibfnamefont{X.}~\bibnamefont{Liu}},
  \bibinfo{author}{\bibfnamefont{F.}~\bibnamefont{Yuan}}, \bibnamefont{and}
  \bibinfo{author}{\bibfnamefont{H.~X.} \bibnamefont{Zhu}},
  \bibinfo{journal}{Research} \textbf{\bibinfo{volume}{2025}},
  \bibinfo{pages}{0552} (\bibinfo{year}{2025}{\natexlab{a}}),
  \eprint{2406.05880}.

\bibitem[{\citenamefont{Guo et~al.}(2025{\natexlab{b}})\citenamefont{Guo, Liu,
  and Yuan}}]{Guo:2024vpe}
\bibinfo{author}{\bibfnamefont{Y.}~\bibnamefont{Guo}},
  \bibinfo{author}{\bibfnamefont{X.}~\bibnamefont{Liu}}, \bibnamefont{and}
  \bibinfo{author}{\bibfnamefont{F.}~\bibnamefont{Yuan}},
  \bibinfo{journal}{Phys. Rev. D} \textbf{\bibinfo{volume}{112}},
  \bibinfo{pages}{054006} (\bibinfo{year}{2025}{\natexlab{b}}),
  \eprint{2408.14693}.

\bibitem[{sm()}]{sm}
\bibinfo{note}{See Supplemental Material for the relations of FECs with DiFFs
  and EECs, details of the fitting and modeling procedure of FECs, and the hard
  coefficients $(C_{abc}, T_{abc})$ in Eq.~(12).}

\bibitem[{\citenamefont{Bierlich et~al.}(2022)}]{Bierlich:2022pfr}
\bibinfo{author}{\bibfnamefont{C.}~\bibnamefont{Bierlich}}
  \bibnamefont{et~al.}, \bibinfo{journal}{SciPost Phys. Codeb.}
  \textbf{\bibinfo{volume}{2022}}, \bibinfo{pages}{8} (\bibinfo{year}{2022}),
  \eprint{2203.11601}.

\bibitem[{\citenamefont{Kang et~al.}(2026{\natexlab{b}})\citenamefont{Kang,
  Metz, Pitonyak, and Zhang}}]{Kang:2026qbm}
\bibinfo{author}{\bibfnamefont{Z.-B.} \bibnamefont{Kang}},
  \bibinfo{author}{\bibfnamefont{A.}~\bibnamefont{Metz}},
  \bibinfo{author}{\bibfnamefont{D.}~\bibnamefont{Pitonyak}}, \bibnamefont{and}
  \bibinfo{author}{\bibfnamefont{C.}~\bibnamefont{Zhang}},
  \bibinfo{journal}{Phys. Lett. B} \textbf{\bibinfo{volume}{882}},
  \bibinfo{pages}{141004} (\bibinfo{year}{2026}{\natexlab{b}}),
  \eprint{2607.02441}.

\bibitem[{\citenamefont{Bacchetta et~al.}(2008)\citenamefont{Bacchetta, Boer,
  Diehl, and Mulders}}]{Bacchetta:2008xw}
\bibinfo{author}{\bibfnamefont{A.}~\bibnamefont{Bacchetta}},
  \bibinfo{author}{\bibfnamefont{D.}~\bibnamefont{Boer}},
  \bibinfo{author}{\bibfnamefont{M.}~\bibnamefont{Diehl}}, \bibnamefont{and}
  \bibinfo{author}{\bibfnamefont{P.~J.} \bibnamefont{Mulders}},
  \bibinfo{journal}{JHEP} \textbf{\bibinfo{volume}{08}}, \bibinfo{pages}{023}
  (\bibinfo{year}{2008}), \eprint{0803.0227}.

\bibitem[{\citenamefont{Yuan and Zhou}(2009)}]{Yuan:2009dw}
\bibinfo{author}{\bibfnamefont{F.}~\bibnamefont{Yuan}} \bibnamefont{and}
  \bibinfo{author}{\bibfnamefont{J.}~\bibnamefont{Zhou}},
  \bibinfo{journal}{Phys. Rev. Lett.} \textbf{\bibinfo{volume}{103}},
  \bibinfo{pages}{052001} (\bibinfo{year}{2009}), \eprint{0903.4680}.

\bibitem[{\citenamefont{Zhou and Metz}(2011)}]{Zhou:2011ba}
\bibinfo{author}{\bibfnamefont{J.}~\bibnamefont{Zhou}} \bibnamefont{and}
  \bibinfo{author}{\bibfnamefont{A.}~\bibnamefont{Metz}},
  \bibinfo{journal}{Phys. Rev. Lett.} \textbf{\bibinfo{volume}{106}},
  \bibinfo{pages}{172001} (\bibinfo{year}{2011}), \eprint{1101.3273}.

\bibitem[{\citenamefont{Gamberg et~al.}(2022)\citenamefont{Gamberg, Malda,
  Miller, Pitonyak, Prokudin, and Sato}}]{Gamberg:2022kdb}
\bibinfo{author}{\bibfnamefont{L.}~\bibnamefont{Gamberg}},
  \bibinfo{author}{\bibfnamefont{M.}~\bibnamefont{Malda}},
  \bibinfo{author}{\bibfnamefont{J.~A.} \bibnamefont{Miller}},
  \bibinfo{author}{\bibfnamefont{D.}~\bibnamefont{Pitonyak}},
  \bibinfo{author}{\bibfnamefont{A.}~\bibnamefont{Prokudin}}, \bibnamefont{and}
  \bibinfo{author}{\bibfnamefont{N.}~\bibnamefont{Sato}}
  (\bibinfo{collaboration}{Jefferson Lab Angular Momentum (JAM), Jefferson Lab
  Angular Momentum}), \bibinfo{journal}{Phys. Rev. D}
  \textbf{\bibinfo{volume}{106}}, \bibinfo{pages}{034014}
  (\bibinfo{year}{2022}), \eprint{2205.00999}.

\bibitem[{\citenamefont{Nayak et~al.}(2005)\citenamefont{Nayak, Qiu, and
  Sterman}}]{Nayak:2005rt}
\bibinfo{author}{\bibfnamefont{G.~C.} \bibnamefont{Nayak}},
  \bibinfo{author}{\bibfnamefont{J.-W.} \bibnamefont{Qiu}}, \bibnamefont{and}
  \bibinfo{author}{\bibfnamefont{G.~F.} \bibnamefont{Sterman}},
  \bibinfo{journal}{Phys. Rev. D} \textbf{\bibinfo{volume}{72}},
  \bibinfo{pages}{114012} (\bibinfo{year}{2005}), \eprint{hep-ph/0509021}.

\bibitem[{\citenamefont{Bodwin}(1985)}]{Bodwin:1984hc}
\bibinfo{author}{\bibfnamefont{G.~T.} \bibnamefont{Bodwin}},
  \bibinfo{journal}{Phys. Rev. D} \textbf{\bibinfo{volume}{31}},
  \bibinfo{pages}{2616} (\bibinfo{year}{1985}), \bibinfo{note}{[Erratum:
  Phys.Rev.D 34, 3932 (1986)]}.

\bibitem[{\citenamefont{Collins et~al.}(1985)\citenamefont{Collins, Soper, and
  Sterman}}]{Collins:1985ue}
\bibinfo{author}{\bibfnamefont{J.~C.} \bibnamefont{Collins}},
  \bibinfo{author}{\bibfnamefont{D.~E.} \bibnamefont{Soper}}, \bibnamefont{and}
  \bibinfo{author}{\bibfnamefont{G.~F.} \bibnamefont{Sterman}},
  \bibinfo{journal}{Nucl. Phys. B} \textbf{\bibinfo{volume}{261}},
  \bibinfo{pages}{104} (\bibinfo{year}{1985}).

\bibitem[{\citenamefont{Collins et~al.}(1988)\citenamefont{Collins, Soper, and
  Sterman}}]{Collins:1988ig}
\bibinfo{author}{\bibfnamefont{J.~C.} \bibnamefont{Collins}},
  \bibinfo{author}{\bibfnamefont{D.~E.} \bibnamefont{Soper}}, \bibnamefont{and}
  \bibinfo{author}{\bibfnamefont{G.~F.} \bibnamefont{Sterman}},
  \bibinfo{journal}{Nucl. Phys. B} \textbf{\bibinfo{volume}{308}},
  \bibinfo{pages}{833} (\bibinfo{year}{1988}).

\bibitem[{\citenamefont{Collins}(2011)}]{Collins:2011zzd}
\bibinfo{author}{\bibfnamefont{J.}~\bibnamefont{Collins}},
  \emph{\bibinfo{title}{{Foundations of Perturbative QCD}}},
  vol.~\bibinfo{volume}{32} (\bibinfo{publisher}{Cambridge University Press},
  \bibinfo{year}{2011}), ISBN \bibinfo{isbn}{978-1-009-40184-5,
  978-1-009-40183-8, 978-1-009-40182-1}.

\bibitem[{\citenamefont{Abe et~al.}(2006)}]{Belle:2005dmx}
\bibinfo{author}{\bibfnamefont{K.}~\bibnamefont{Abe}} \bibnamefont{et~al.}
  (\bibinfo{collaboration}{Belle}), \bibinfo{journal}{Phys. Rev. Lett.}
  \textbf{\bibinfo{volume}{96}}, \bibinfo{pages}{232002}
  (\bibinfo{year}{2006}), \eprint{hep-ex/0507063}.

\bibitem[{\citenamefont{Seidl et~al.}(2008)}]{Belle:2008fdv}
\bibinfo{author}{\bibfnamefont{R.}~\bibnamefont{Seidl}} \bibnamefont{et~al.}
  (\bibinfo{collaboration}{Belle}), \bibinfo{journal}{Phys. Rev. D}
  \textbf{\bibinfo{volume}{78}}, \bibinfo{pages}{032011}
  (\bibinfo{year}{2008}), \bibinfo{note}{[Erratum: Phys.Rev.D 86, 039905
  (2012)]}, \eprint{0805.2975}.

\bibitem[{\citenamefont{Lees et~al.}(2014)}]{BaBar:2013jdt}
\bibinfo{author}{\bibfnamefont{J.~P.} \bibnamefont{Lees}} \bibnamefont{et~al.}
  (\bibinfo{collaboration}{BaBar}), \bibinfo{journal}{Phys. Rev. D}
  \textbf{\bibinfo{volume}{90}}, \bibinfo{pages}{052003}
  (\bibinfo{year}{2014}), \eprint{1309.5278}.

\bibitem[{\citenamefont{Airapetian et~al.}(2005)}]{HERMES:2004mhh}
\bibinfo{author}{\bibfnamefont{A.}~\bibnamefont{Airapetian}}
  \bibnamefont{et~al.} (\bibinfo{collaboration}{HERMES}),
  \bibinfo{journal}{Phys. Rev. Lett.} \textbf{\bibinfo{volume}{94}},
  \bibinfo{pages}{012002} (\bibinfo{year}{2005}), \eprint{hep-ex/0408013}.

\bibitem[{\citenamefont{Airapetian et~al.}(2010)}]{HERMES:2010mmo}
\bibinfo{author}{\bibfnamefont{A.}~\bibnamefont{Airapetian}}
  \bibnamefont{et~al.} (\bibinfo{collaboration}{HERMES}),
  \bibinfo{journal}{Phys. Lett. B} \textbf{\bibinfo{volume}{693}},
  \bibinfo{pages}{11} (\bibinfo{year}{2010}), \eprint{1006.4221}.

\bibitem[{\citenamefont{Adolph et~al.}(2012)}]{COMPASS:2012ozz}
\bibinfo{author}{\bibfnamefont{C.}~\bibnamefont{Adolph}} \bibnamefont{et~al.}
  (\bibinfo{collaboration}{COMPASS}), \bibinfo{journal}{Phys. Lett. B}
  \textbf{\bibinfo{volume}{717}}, \bibinfo{pages}{376} (\bibinfo{year}{2012}),
  \eprint{1205.5121}.

\bibitem[{\citenamefont{Abdallah et~al.}(2022)}]{STAR:2022hqg}
\bibinfo{author}{\bibfnamefont{M.}~\bibnamefont{Abdallah}} \bibnamefont{et~al.}
  (\bibinfo{collaboration}{STAR}), \bibinfo{journal}{Phys. Rev. D}
  \textbf{\bibinfo{volume}{106}}, \bibinfo{pages}{072010}
  (\bibinfo{year}{2022}), \eprint{2205.11800}.

\bibitem[{\citenamefont{Aboona et~al.}(2025{\natexlab{b}})}]{STAR:2025xyp}
\bibinfo{author}{\bibfnamefont{B.}~\bibnamefont{Aboona}} \bibnamefont{et~al.}
  (\bibinfo{collaboration}{STAR}), \bibinfo{journal}{Phys. Rev. Lett.}
  \textbf{\bibinfo{volume}{135}}, \bibinfo{pages}{261902}
  (\bibinfo{year}{2025}{\natexlab{b}}), \eprint{2507.16355}.

\bibitem[{\citenamefont{Aboona et~al.}(2026)}]{STARCollaboration:2026kaa}
\bibinfo{author}{\bibfnamefont{B.~E.} \bibnamefont{Aboona}}
  \bibnamefont{et~al.} (\bibinfo{collaboration}{(STAR Collaboration)*, STAR}),
  \bibinfo{journal}{Phys. Rev. Lett.} \textbf{\bibinfo{volume}{137}},
  \bibinfo{pages}{111902} (\bibinfo{year}{2026}), \eprint{2604.15543}.

\bibitem[{\citenamefont{Artru and Collins}(1996)}]{Artru:1995zu}
\bibinfo{author}{\bibfnamefont{X.}~\bibnamefont{Artru}} \bibnamefont{and}
  \bibinfo{author}{\bibfnamefont{J.~C.} \bibnamefont{Collins}},
  \bibinfo{journal}{Z. Phys. C} \textbf{\bibinfo{volume}{69}},
  \bibinfo{pages}{277} (\bibinfo{year}{1996}), \eprint{hep-ph/9504220}.

\bibitem[{\citenamefont{Boer et~al.}(2003)\citenamefont{Boer, Jakob, and
  Radici}}]{Boer:2003ya}
\bibinfo{author}{\bibfnamefont{D.}~\bibnamefont{Boer}},
  \bibinfo{author}{\bibfnamefont{R.}~\bibnamefont{Jakob}}, \bibnamefont{and}
  \bibinfo{author}{\bibfnamefont{M.}~\bibnamefont{Radici}},
  \bibinfo{journal}{Phys. Rev. D} \textbf{\bibinfo{volume}{67}},
  \bibinfo{pages}{094003} (\bibinfo{year}{2003}), \bibinfo{note}{[Erratum:
  Phys.Rev.D 98, 039902 (2018)]}, \eprint{hep-ph/0302232}.

\bibitem[{\citenamefont{Vossen et~al.}(2011)}]{Belle:2011cur}
\bibinfo{author}{\bibfnamefont{A.}~\bibnamefont{Vossen}} \bibnamefont{et~al.}
  (\bibinfo{collaboration}{Belle}), \bibinfo{journal}{Phys. Rev. Lett.}
  \textbf{\bibinfo{volume}{107}}, \bibinfo{pages}{072004}
  (\bibinfo{year}{2011}), \eprint{1104.2425}.

\bibitem[{\citenamefont{Cheng and Yan}(2025)}]{Cheng:2025cuv}
\bibinfo{author}{\bibfnamefont{K.}~\bibnamefont{Cheng}} \bibnamefont{and}
  \bibinfo{author}{\bibfnamefont{B.}~\bibnamefont{Yan}},
  \bibinfo{journal}{Phys. Rev. Lett.} \textbf{\bibinfo{volume}{135}},
  \bibinfo{pages}{011902} (\bibinfo{year}{2025}), \eprint{2501.03321}.

\bibitem[{\citenamefont{Liu et~al.}(2026)\citenamefont{Liu, Qi, He, and
  Xiao}}]{Liu:2026dzv}
\bibinfo{author}{\bibfnamefont{Y.-X.} \bibnamefont{Liu}},
  \bibinfo{author}{\bibfnamefont{W.}~\bibnamefont{Qi}},
  \bibinfo{author}{\bibfnamefont{L.-T.} \bibnamefont{He}}, \bibnamefont{and}
  \bibinfo{author}{\bibfnamefont{B.-W.} \bibnamefont{Xiao}}
  (\bibinfo{year}{2026}), \eprint{2604.17756}.

\bibitem[{\citenamefont{Fucilla et~al.}(2026)\citenamefont{Fucilla, Guo, and
  Hatta}}]{Fucilla:2026zhy}
\bibinfo{author}{\bibfnamefont{M.}~\bibnamefont{Fucilla}},
  \bibinfo{author}{\bibfnamefont{Y.}~\bibnamefont{Guo}}, \bibnamefont{and}
  \bibinfo{author}{\bibfnamefont{Y.}~\bibnamefont{Hatta}}
  (\bibinfo{year}{2026}), \eprint{2609.32800}.

\bibitem[{\citenamefont{Pitonyak et~al.}(2024)\citenamefont{Pitonyak, Cocuzza,
  Metz, Prokudin, and Sato}}]{Pitonyak:2023gjx}
\bibinfo{author}{\bibfnamefont{D.}~\bibnamefont{Pitonyak}},
  \bibinfo{author}{\bibfnamefont{C.}~\bibnamefont{Cocuzza}},
  \bibinfo{author}{\bibfnamefont{A.}~\bibnamefont{Metz}},
  \bibinfo{author}{\bibfnamefont{A.}~\bibnamefont{Prokudin}}, \bibnamefont{and}
  \bibinfo{author}{\bibfnamefont{N.}~\bibnamefont{Sato}},
  \bibinfo{journal}{Phys. Rev. Lett.} \textbf{\bibinfo{volume}{132}},
  \bibinfo{pages}{011902} (\bibinfo{year}{2024}), \eprint{2305.11995}.

\bibitem[{\citenamefont{Cocuzza et~al.}(2024)\citenamefont{Cocuzza, Metz,
  Pitonyak, Prokudin, Sato, and Seidl}}]{Cocuzza:2023vqs}
\bibinfo{author}{\bibfnamefont{C.}~\bibnamefont{Cocuzza}},
  \bibinfo{author}{\bibfnamefont{A.}~\bibnamefont{Metz}},
  \bibinfo{author}{\bibfnamefont{D.}~\bibnamefont{Pitonyak}},
  \bibinfo{author}{\bibfnamefont{A.}~\bibnamefont{Prokudin}},
  \bibinfo{author}{\bibfnamefont{N.}~\bibnamefont{Sato}}, \bibnamefont{and}
  \bibinfo{author}{\bibfnamefont{R.}~\bibnamefont{Seidl}}
  (\bibinfo{collaboration}{Jefferson Lab Angular Momentum (JAM)}),
  \bibinfo{journal}{Phys. Rev. D} \textbf{\bibinfo{volume}{109}},
  \bibinfo{pages}{034024} (\bibinfo{year}{2024}), \eprint{2308.14857}.

\bibitem[{\citenamefont{Pitonyak et~al.}(2026)\citenamefont{Pitonyak, Cocuzza,
  Metz, Prokudin, and Sato}}]{Pitonyak:2025lin}
\bibinfo{author}{\bibfnamefont{D.}~\bibnamefont{Pitonyak}},
  \bibinfo{author}{\bibfnamefont{C.}~\bibnamefont{Cocuzza}},
  \bibinfo{author}{\bibfnamefont{A.}~\bibnamefont{Metz}},
  \bibinfo{author}{\bibfnamefont{A.}~\bibnamefont{Prokudin}}, \bibnamefont{and}
  \bibinfo{author}{\bibfnamefont{N.}~\bibnamefont{Sato}},
  \bibinfo{journal}{Phys. Rev. D} \textbf{\bibinfo{volume}{113}},
  \bibinfo{pages}{038901} (\bibinfo{year}{2026}), \eprint{2502.15817}.

\bibitem[{\citenamefont{Kang et~al.}(2024)\citenamefont{Kang, Lee, Shao, and
  Zhao}}]{Kang:2023big}
\bibinfo{author}{\bibfnamefont{Z.-B.} \bibnamefont{Kang}},
  \bibinfo{author}{\bibfnamefont{K.}~\bibnamefont{Lee}},
  \bibinfo{author}{\bibfnamefont{D.~Y.} \bibnamefont{Shao}}, \bibnamefont{and}
  \bibinfo{author}{\bibfnamefont{F.}~\bibnamefont{Zhao}},
  \bibinfo{journal}{JHEP} \textbf{\bibinfo{volume}{03}}, \bibinfo{pages}{153}
  (\bibinfo{year}{2024}), \eprint{2310.15159}.

\bibitem[{\citenamefont{Kang et~al.}(2026{\natexlab{c}})\citenamefont{Kang,
  Metz, Pitonyak, and Zhang}}]{Kang:2026pro}
\bibinfo{author}{\bibfnamefont{Z.-B.} \bibnamefont{Kang}},
  \bibinfo{author}{\bibfnamefont{A.}~\bibnamefont{Metz}},
  \bibinfo{author}{\bibfnamefont{D.}~\bibnamefont{Pitonyak}}, \bibnamefont{and}
  \bibinfo{author}{\bibfnamefont{C.}~\bibnamefont{Zhang}}
  (\bibinfo{year}{2026}{\natexlab{c}}), \eprint{2604.28131}.

\bibitem[{\citenamefont{Altarelli and Parisi}(1977)}]{Altarelli:1977zs}
\bibinfo{author}{\bibfnamefont{G.}~\bibnamefont{Altarelli}} \bibnamefont{and}
  \bibinfo{author}{\bibfnamefont{G.}~\bibnamefont{Parisi}},
  \bibinfo{journal}{Nucl. Phys. B} \textbf{\bibinfo{volume}{126}},
  \bibinfo{pages}{298} (\bibinfo{year}{1977}).

\bibitem[{\citenamefont{Artru and Mekhfi}(1990)}]{Artru:1989zv}
\bibinfo{author}{\bibfnamefont{X.}~\bibnamefont{Artru}} \bibnamefont{and}
  \bibinfo{author}{\bibfnamefont{M.}~\bibnamefont{Mekhfi}},
  \bibinfo{journal}{Z. Phys. C} \textbf{\bibinfo{volume}{45}},
  \bibinfo{pages}{669} (\bibinfo{year}{1990}).

\bibitem[{\citenamefont{Stratmann and Vogelsang}(2002)}]{Stratmann:2001pt}
\bibinfo{author}{\bibfnamefont{M.}~\bibnamefont{Stratmann}} \bibnamefont{and}
  \bibinfo{author}{\bibfnamefont{W.}~\bibnamefont{Vogelsang}},
  \bibinfo{journal}{Phys. Rev. D} \textbf{\bibinfo{volume}{65}},
  \bibinfo{pages}{057502} (\bibinfo{year}{2002}), \eprint{hep-ph/0108241}.

\bibitem[{\citenamefont{Bomhof and Mulders}(2007)}]{Bomhof:2006ra}
\bibinfo{author}{\bibfnamefont{C.~J.} \bibnamefont{Bomhof}} \bibnamefont{and}
  \bibinfo{author}{\bibfnamefont{P.~J.} \bibnamefont{Mulders}},
  \bibinfo{journal}{JHEP} \textbf{\bibinfo{volume}{02}}, \bibinfo{pages}{029}
  (\bibinfo{year}{2007}), \eprint{hep-ph/0609206}.

\bibitem[{\citenamefont{Yuan}(2008{\natexlab{b}})}]{Yuan:2007nd}
\bibinfo{author}{\bibfnamefont{F.}~\bibnamefont{Yuan}}, \bibinfo{journal}{Phys.
  Rev. Lett.} \textbf{\bibinfo{volume}{100}}, \bibinfo{pages}{032003}
  (\bibinfo{year}{2008}{\natexlab{b}}), \eprint{0709.3272}.

\end{thebibliography}

\clearpage
\onecolumngrid
\raggedbottom
\renewcommand{\topfraction}{0.90}
\renewcommand{\bottomfraction}{0.85}
\renewcommand{\textfraction}{0.10}
\renewcommand{\floatpagefraction}{0.75}
\setcounter{topnumber}{4}
\setcounter{bottomnumber}{3}
\setcounter{totalnumber}{6}
\setlength{\floatsep}{12pt plus 2pt minus 2pt}
\setlength{\textfloatsep}{12pt plus 2pt minus 2pt}
\setlength{\intextsep}{12pt plus 2pt minus 2pt}
\makeatletter
\setlength{\@fptop}{0pt}
\setlength{\@fpsep}{12pt plus 2pt minus 2pt}
\setlength{\@fpbot}{0pt plus 1fil}
\makeatother
\appendix
\makeatletter
\@removefromreset{equation}{section}
\def\@sectioncntformat#1{\csname the#1\endcsname}
\makeatother
\setcounter{equation}{0}
\setcounter{figure}{0}
\setcounter{table}{0}
\setcounter{page}{1}
\renewcommand{\theequation}{S\arabic{equation}}
\renewcommand{\thefigure}{S\arabic{figure}}
\renewcommand{\thetable}{S\arabic{table}}

\begingroup
\makeatletter
\def\@title{Supplemental material for ``\sm@title''}
\let\@AAC@list\sm@authors
\let\@AFF@list\sm@affiliations
\frontmatter@setup
\let\@makefnmark\frontmatter@makefnmark
\thispagestyle{titlepage}
\frontmatter@title@produce
\frontmatter@author@produce@script
\frontmatter@RRAPformat{\expandafter\produce@RRAP\expandafter{\@date}}
\par\frontmatter@finalspace
\makeatother
\endgroup

\section{Relation of FEC to DiFF and EEC}
\label{app:fec-diff}
In this section, we derive the relation of FECs to DiFFs and EECs.
With one tagged hadron and one energy-flow measurement, the FEC provides an intermediate description between a DiFF, which measures two explicit hadrons, and an EEC, which measures two energy flows.
The relation of the latter two has been established for the understanding of the collinear limit of EEC~\cite{Lee:2025okn, Kang:2025zto, Chang:2025kgq, Guo:2025zwb, Kang:2026qbm}.

We begin by restoring the Wilson lines suppressed in the FEC quark correlator in \eq{eq:fec-matrix} of the main text,
\begin{align}
  \Delta^{\alpha\beta}_{h/q}(z, \qf, \phi)
  = \frac{z}{4N_c} \sum_X
	&\int \frac{dw^-}{2\pi p_h^+}\,
	e^{i p_h^+ w^-/z} \, \langle 0| \, W(\infty, w^-)\, \psi_{q}^{\alpha}(w^-)\,
	 \E_L(\qf,\phi) \, |h,X\rangle \,
	\langle h,X| \, \bar{\psi}_{q}^{\beta}(0) \, W^\dagger(\infty,0)\,|0\rangle
	\,,
\label{eq:fec-matrix-full}
\end{align}
where the spinor indices $\alpha$ and $\beta$ are left open.
The observed hadron $h$ is taken to 
move along the $\hat z_h$ direction,
with respect to which we define a lightfront coordinate system.
The corresponding lightlike basis vectors have the Cartesian components $\bar n^{\mu} = (1,0,0,1)/\sqrt{2}$ and $n^{\mu} = (1,0,0,-1)/\sqrt{2}$ pointing along and opposite to the $\hat z_h$ direction, respectively. 
Then any vector $v^{\mu}$ is specified by
$$
v^{\mu} = (v^+, v^-, \bm{v}_T) \equiv \pp{v\cdot n, v\cdot \bar n, \bm v_T},
$$
with $v^{\pm} \equiv (v^0 \pm v^3) / \sqrt{2}$ 
and $\bm{v}_T \equiv (v^1, v^2)$.
The two quark fields are located on the lightcone in the direction opposite to the $h$ momentum $p_h$,
separated by a distance $w^-$
that is integrated under a Fourier transformation, 
fixing the quark's longitudinal momentum to be $k^+ = p_h^+ / z$.
The two future-pointing Wilson lines in the fundamental representation connect the quark fields through light-cone infinity, ensuring gauge invariance. They are defined for a generic starting point $x$ as
\begin{equation}
    W(\infty, x) = \mathcal{P} \exp\bb{ -ig\int_{0}^{\infty} d\lambda\, A^+(x + \lambda n)
    },
    \label{eq:app-wilson-line}
\end{equation}
where $A^+ \equiv n_{\mu} A^{a \mu} \, t^a$ with $t^a$ being the color generators, and $\mathcal{P}$ denotes path ordering.

The DiFF correlator has the same quark field and Wilson line structure,
but identifies two explicit hadrons $(h', h)$.
It is defined as~\cite{Pitonyak:2023gjx, Cocuzza:2023vqs, Pitonyak:2025lin}
\begin{align}
	\Psi^{\alpha\beta}_{h' h / q}(z', z, \bm{R}_T)
	&= \frac{(z' + z)^2}{64 \pi^3 z' z N_c} 
		\sum_X \int \frac{dw^-}{2\pi} \, e^{i k^+ w^-} \,
		\vv{ 0 | W(\infty, w^-)\,  \psi_q^{\alpha}(w^-) \, | h'(p_{h'}), h(p_h), X }
	\nn\\
	&\hspace{13.6em} \times 
			\vv{ h'(p_{h'}), h(p_h), X | \, \bar{\psi}_q^{\beta}(0) \, W^\dagger(\infty,0) \, | 0 },
\label{eq:diff-def}
\end{align}
where the $\hat{z}_{hh}$ direction is along the dihadron momentum
$P_{hh} = p_{h'} + p_h$, with respect to which the DiFF lightfront coordinates are defined,
with $z' = p_{h'}^+ / k^+$ and $z = p_h^+ / k^+$.
The $\bm{R}_T = (\bm{p}_{h' T} - \bm{p}_{hT}) / 2 = R_T \, (\cos\phi_R, \sin\phi_R)$ is half of the relative transverse momentum between $h'$ and $h$ in this frame.
Similar to \eq{eq:fec-spin}, at leading twist, \eq{eq:diff-def} also gives two scalar DiFFs,
\begin{align*}
  	{\rm Tr} \!\left[ \gamma^+ \, \Psi_{h' h / q} \right]
  &= D_{h' h / q}(z', z, R_T^2), 
  \nn\\
  	{\rm Tr} \!\left[ \gamma^+\gamma_\perp^i\gamma_5 \, \Psi_{h' h / q} \right]
  &= H_{h' h / q}(z', z, R_T^2) \, \frac{\pp{ \hat{z}_{hh} \times \bm{R}_{T} }^i }{M_{hh}}
  \,,
\end{align*}
where $D_{h' h / q}$ and $H_{h' h / q}$ describe
unpolarized and transversely polarized quark fragmentation, respectively,
and 
$M_{hh}^2 = P_{hh}^2 = (z' + z)^2 R_T^2 / (z' z)$
is the invariant mass squared of the dihadron system in the massless limit.

Performing a Lorentz transformation to reorient 
the $\hat z_{hh}$ direction in \eq{eq:diff-def} along $p_h$
and summing over the other hadron $h'$ with a weight of its plus momentum $p_{h'}^+$,
we can relate it to the FEC in \eq{eq:fec-matrix-full},
\begin{align}
	&\frac{\Delta^{\alpha\beta}_{h/q}(z, \qf, \phi)}{2\qf^2}
	= \sum_{h'} \int_0^{1-z} dz' \frac{z'}{z} \int d^2\bm{R}_{T} \, 
		\frac{R_T^2}{\qf^2} \,
		\delta(\phi - \phi_R) \,
		\delta\pp{ R_T^2 - \frac{z'^2 \qf^2}{(z' + z)^2} } \,
		\bb{
			\Psi^{\alpha\beta}_{h' h / q}\pp{z', z, \bm{R}_T }
		}.
\label{eq:fec-diff}
\end{align}
Similar relations hold between $(\unpfec, \colfec)$ and $(D,H)$.
These relations are consistent
with the universality of both FECs and DiFFs
and identify the role of $\qf$ as the analogue of $R_T$ (or $M_{hh}$) in DiFFs,
$\qf^2 = (z' + z)^2 R_T^2 / z'^2$.

The major advantage of FEC over DiFF is that,
with $z'$ and $R_T$ integrated,
FEC alleviates the dependence on resonance states suffered by DiFFs~\cite{Cocuzza:2023vqs}.
This makes $\qf$ a better scale scanner than $R_T$ or $M_{hh}$.

As for DiFFs~\cite{Pitonyak:2023gjx}, we can obtain a momentum conservation sum rule that relates the unpolarized FEC to the ordinary single hadron fragmentation function $d_{h/q}(z)$.
This can be readily seen from the definition in \eq{eq:fec-matrix-full}.
Since the tagged hadron $h$ carries a longitudinal momentum $p_h^+ = z k^+$ out of the fragmenting parton of momentum $k$, the rest of the fragmentation products, $X$, must carry a total longitudinal momentum
$p_X^+ = \sum_{j\in X} p_j^+ = (1 - z) k^+$.
By integrating over $\Lambda_F^2$ and $\phi$ in \eq{eq:fec-matrix-full}, or, more specifically, for the $\unpfec$ in \eq{eq:fec-def-D}, 
the energy flow simply turns into a weight of $p_X^+ / p_h^+$ acting on the same final states that are involved in the definition of unpolarized fragmentation function $d(z)$ 
and gives
\begin{equation}
    2\pi \int_0^\infty d\Lambda_F^2\,
    \mathcal D_{h/q}(z,\Lambda_F^2)  = \frac{1-z}{z}\,d_{h/q}(z).
\label{eq:fec-sum-rule}
\end{equation}
By virtue of the use of $p_j^+$, instead of energy, as the weight in $\E_L$, this relation is an exact result.

Further summing over $h$ in \eq{eq:fec-diff} 
along with its energy weight will relate FEC to EEC.
Interestingly, because EEC is symmetric with its two energy flows,
the azimuthal asymmetry vanishes automatically.
Only one of the two relations survives,
\begin{align}
	D^{\rm EEC}_i(\chi)
    &= \sum_{h', h} \int dz' \, dz \, d^2 \bm{R}_T \,
    	\delta\pp{ \chi - \frac{(z' + z)^2 R_T^2}{z'^2 z^2} }
        \,
    	z' z \, D_1^{h' h / i}(z', z, R^2_T)\nn\\
	&= 2\pi \sum_h \int_0^1 dz \, z^4 \, 
		\unpfec_{h/i}(z, \qf^2 = z^2 \chi) \,,
\label{eq:fec-eec}
\end{align}
where the left-hand side is the EEC-DiFF defined with the DiFF in the first line~\cite{Kang:2025zto} (we have introduced an extra $1/s$ factor compared to \refcite{Kang:2025zto}).
The $\chi$ is the scaling variable in EEC, 
approximating to $s \, \theta^2 / 4$ for small EEC angle $\theta$
in $e^+e^-$ collisions at c.m.\ energy $\sqrt{s}$.
The $z^4$ weight in the second line comes from 
(1) a $z$ factor for the energy weight of $h$,
(2) a $z^2$ for compensating the normalization of $\mathcal{D}_i(\chi)$ in terms of unit $\chi$
	and $\unpfec_{h/i}(z, \qf^2)$ in terms of unit $\qf^2$,
and
(3) another $z$ factor due to the factor $1 / p_h^+$ in \eq{eq:fec-matrix} of the main text.

We note that \refs{Kang:2023big, Kang:2026pro, Kang:2026qbm}
restore the Collins asymmetry from EEC 
by restricting the hadron types
measured in both of the two energy flows.
In comparison, FEC achieves this 
with the explicit hadron $h$ by definition. 
One could in principle also impose a quantum number restriction to the energy flow in FEC 
such as charged tracks~\cite{Chang:2013rca,Chang:2013iba};
this extension is not pursued in this work.

One important implication of \eq{eq:fec-eec} is that 
the EEC scaling variable $\chi$ corresponds to 
the scale $\qf^2 / z^2$ in the FEC.
As emphasized in the main text, 
this provides a key insight for the phenomenological modeling of FEC.

Lastly, we point out that as $\qf \gg \Lambda_{\rm QCD}$, the FECs themselves can be factorized into inclusive collinear fragmentation functions, with perturbatively calculable matching coefficients that have characteristic hard scale at $\qf$.
In fact, for the unpolarized FEC $\unpfec$, this matching has been given in \refcite{Zhu:2025qkx}.
Adapting it to our notations, we have
\begin{equation}
    \qf^2 \, \unpfec_{h/q}(z, \qf^2)
    = \frac{\alpha_s(\qf) \, C_F}{4 \pi^2}
    \int_z^1 \frac{d\xi}{\xi^2} \,
    \biggcc{
        \frac{1+u^2}{u} \,
        d_{h/q}(\xi)
        +
        \frac{(1-u) \bb{1+(1-u)^2}}{u^2}\,
        d_{h/g}(\xi)
    }
    + \mathcal{O}\pp{ \frac{\Lambda_{\rm QCD}}{\qf}, \alpha_s^2 },
\label{eq:fec-ope-matching}
\end{equation}
where $u \equiv z / \xi$
and we have neglected the dependence on the renormalization scale at LO.
Similarly, the matching of the Collins FEC at large $\qf$ can be obtained following that of the TMD Collins fragmentation function at large transverse momentum~\cite{Yuan:2009dw}.
Its leading power, however, starts at twist three,
\begin{equation}
    \qf^3 \, \colfec_{h/q}(z, \qf^2)
    =
    \frac{\alpha_s(\qf) \, C_F}{4\pi^2}
    \int_z^1 \frac{d\xi}{\xi^2}
    \bb{
    2\hat H(\xi)
    +\dots
    }
    + \mathcal{O}\pp{ \frac{\Lambda_{\rm QCD}}{\qf}, \alpha_s^2 },
\label{eq:fec-ope-matching1}
\end{equation}
where $\hat H(\xi)$ is the twist-3 fragmentation function related to the $k_T$ moment of the Collins function on the right-hand side of \eq{eq:moment-relation}.
The ellipsis denotes contributions from other twist-3 channels,
and we leave the detailed calculation to a future work. 

\section{FEC modeling, evolution, and universality check}
\label{app:fec-modeling}

As stated in the main text, we parametrize the FECs by \eqs{eq:fec-param-D}{eq:fec-param-H} at the initial scale $\mu_0 = 5~\mathrm{GeV}$
and, for any given values of the parameters therein, evolve them to other scales by DGLAP equations to make predictions and/or compare with ``data.''
The unpolarized FEC $\unpfec_{h/i}(z, \qf^2; \mu)$ and the Collins FEC $\colfec_{h/i}(z, \qf^2; \mu)$ follow the same DGLAP equations as the unpolarized and the transversity fragmentation functions, respectively,
\begin{align}
    \frac{d \, \unpfec_{h/i}(z,\qf^2;\mu)}
       {d\ln\mu^2}
    &= \sum_{j = q, \bar q, g} \int_z^1 \frac{dw}{w} \,
    \unpfec_{h/j}\pp{ \frac{z}{w},\qf^2;\mu } 
    \, P^T_{ji}(w, \alpha_s(\mu))
    \, ,
    \qquad
    (i = q, \bar q, g)
    \nn\\
    \frac{d \, \colfec_{h/i}(z,\qf^2;\mu)}
       {d\ln\mu^2}
    &= \sum_{j = q, \bar q} \int_z^1 \frac{dw}{w} \,
    \colfec_{h/j}\pp{ \frac{z}{w},\qf^2;\mu } 
    \, \delta P^T_{ji}(w, \alpha_s(\mu))
    \, ,
    \qquad
    (i = q, \bar q)
\label{eq:fec-dglap}
\end{align}
where $\qf^2$ is simply a spectator variable that does not evolve,
and we note that the linearly polarized gluon FEC $\colfec_{h/g}$ does not mix with quark Collins FECs.
The timelike splitting kernels take the perturbative expansion in $\alpha_s$,
\begin{align}
    P^T_{ji}(w, \alpha_s)
    &= \sum_{n = 0}^{\infty} \pp{ \frac{\alpha_s}{2\pi} }^{n + 1} P^{T,(n)}_{ji}(w),
    \nn\\
    \delta P^T_{ji}(w, \alpha_s)
    &= \sum_{n = 0}^{\infty} \pp{ \frac{\alpha_s}{2\pi} }^{n + 1} \delta P^{T,(n)}_{ji}(w).
\end{align}
We retain only the LO terms in our study, with the coefficients given by~\cite{Altarelli:1977zs}
\begin{align}
    P^{T,(0)}_{qq}(w)
    &= P^{T,(0)}_{\bar q \bar q}(w)
    = C_F \bb{
        \frac{2}{(1-w)_+}
        - 1 - w
        + \frac{3}{2} \delta(1-w) },
    \nn\\
    P^{T,(0)}_{gq}(w)
    &= P^{T,(0)}_{g \bar q}(w)
    = C_F \bb{
        \frac{1+(1-w)^2}{w}
    }, 
    \nn\\
    P^{T,(0)}_{qg}(w)
    &= P^{T,(0)}_{\bar q g}(w)
    = T_F \bb{ 
        w^2+(1-w)^2
    },
    \nn\\
    P^{T,(0)}_{gg}(w)
    &= 2C_A \bb{
        \frac{1}{(1 - w)_+}
        + \frac{1 - w}{w} + w(1-w) - 1
    }
    +\frac{\beta_0}{2}\delta(1-w),
\label{eq:splitting-P}
\end{align}
for the unpolarized kernels,
and~\cite{Artru:1989zv, Stratmann:2001pt} 
\begin{align}
    \delta P^{T,(0)}_{qq}(w)
    =\delta P^{T,(0)}_{\bar q\bar q}(w)
    =C_F\bb{
        \frac{2}{(1-w)_+}-2
        +\frac{3}{2}\delta(1-w)
    },
\label{eq:splitting-tP}
\end{align}
for the transversity kernels,
where the color factors and beta-function coefficient are normalized as
\begin{equation}
    C_F = \frac{N_c^2 - 1}{2 N_c},
    \qquad
    C_A = N_c,
    \qquad 
    T_F = \frac{1}{2},
    \qquad
    \beta_0 = \frac{11 C_A - 4 n_f T_F}{3},
\label{eq:color-fac}
\end{equation}
for $N_c = 3$ colors, 
with $n_f$ being the number of active quark flavors. 
At this order, there is no mixing among quarks of different flavors or between quarks and antiquarks.

The ``data'' for the unpolarized FEC are the FECs extracted from the {\tt Pythia} simulated $e^+ e^- \to hX$ events by \eq{eq:ee-1pI-fec-lo}.
The left-hand side of \eq{eq:ee-1pI-fec-lo} can be obtained as a histogram of $(z, \qf^2)$ by generating events of $e^+e^- \to q\bar{q}$ in {\tt Pythia} at c.m.\ energy $\sqrt{s}$, followed by showering and hadronization. 
For each final-state $h$ in each event, we loop through every hadron $j$ around it and fill the histogram at $(z, \qf) = (2E_h / \sqrt{s}, \Lambda_{F, j})$ with the weight $p_j^+/p_h^+$.
This procedure is the same as experimental analyses,
but in {\tt Pythia}, 
we can further assign each $h$ to either $q$ or $\bar q$ hemisphere 
to determine the flavor channel for the right-hand side of \eq{eq:ee-1pI-fec-lo},
and then obtain a two-dimensional image for each $\unpfec_{h/q}(z, \qf^2)$ corresponding to $\mu = \sqrt{s} / 2$.
The simulated events are generated at five different c.m. energies $\sqrt{s} \in \{ 10, 50, 100, 500, 1000 \}~\GeV$.
At each energy scale, $z$ values are restricted to the ten equally spaced nodes in $[0.125, 0.575]$, and $\qf$ is counted with the upper bound $\qf/z \leq 0.05\sqrt{s}$ and $\qf< 10~\GeV$ for factorization to hold.
We denote the unpolarized FECs extracted from above procedure as $\mathcal{D}^{\rm Pythia}_{h/q}(z, \qf^2; \mu)$.
Apparently, if we only used the unpolarized FEC, we could simply interpolate on these histograms without the need to introduce the parametrization in \eq{eq:fec-param-D} and do the fit.
The reasons for the latter are
\begin{itemize}
\item the fitting parameters for $\unpfec$ can serve as useful reference values for the parameters of $\colfec$, which cannot be extracted from {\tt Pythia};

\item combining the parametrization with the DGLAP evolution 	enables us 
	to make use of the high-$s$ data to constrain the large $\qf$ tail of $\unpfec$
	and relate that to low $\mu$,
	overcoming the factorizability limitation $\qf / z \ll \sqrt{s}$ at low energy $s$; and

\item it allows us to add theoretical constraint such as the sum rule in \eq{eq:fec-sum-rule}.
\end{itemize}
Caution must be made for the last point, however, because \eq{eq:fec-sum-rule} is derived at bare operator level; 
both sides of \eq{eq:fec-sum-rule} require renormalization which will break the exact equality. 
Since the extracted or fitted FECs and FFs correspond to the renormalized functions, 
we modify \eq{eq:fec-sum-rule} by requiring 
\begin{equation}
    R_{f}(z; \mu) 
    = \frac{2\pi \int_0^\infty d\qf^2\,
    \unpfec_{f}(z,\qf^2;\mu)}
    {\frac{1-z}{z}\,
    d^{\rm Pythia}_{f}(z; \mu)}
    \leq 1,
\label{eq:Rf}
\end{equation}
in the fit,
where $f$ labels the possible flavor channel for FECs.
Here, $d^{\rm Pythia}_{f}$ denotes the FF extracted from the same set of {\tt Pythia} simulated events onto the same $z$ and $\mu$ grids.
We obtain it using \eq{eq:ee-1pI-fec-lo} by replacing
$\Sigma_{ee \to h X}$ with the unweighted inclusive cross section $\sigma_{ee \to h X}$,
removing the $\qf^2$ dependence on both sides and the factor $2\pi$ from the right-hand side.
This condition implicitly requires $\gamma_{\unpfec, f}(z) > 0$ for the anomalous dimension in \eq{eq:fec-param-D},
as expected from the resummation of the logarithms $\ln(\qf^2 / \mu^2)$ at large $\qf$ in \eq{eq:fec-ope-matching}~\cite{Zhu:2025qkx}.

To find the independent nonperturbative FECs, we use the SU(2) isospin symmetry and the charge conjugation to reduce the number of possible flavor channels $f$ to nine.
For $\mathcal{F} = \unpfec$ or $\colfec$, we have two FECs for quark flavors $r = s, c, b$, with $\mathcal{F}_{\pi^{\pm} / r} = \mathcal{F}_{\pi^{\mp} / \bar r}$, 
and the flavor-favored and -unfavored FECs for $u$ and $d$ quarks are defined in the main text.
The gluon gives one FEC $\mathcal{F}_g = \mathcal{F}_{\pi^{\pm} / g}$,
completing the set of flavor channels included in the unpolarized FEC fit.

We then fit the parameters in \eq{eq:fec-param-D} by minimizing the loss function,
\begin{equation}
    L = \sum_{f, z, \qf, \mu}
    \biggbb{\ln\frac{\unpfec_f^{\rm fit}(z, \qf^2; \mu)}
    {\unpfec_f^{\rm Pythia}(z, \qf^2; \mu)}}^2
    + w_E \sum_{f, z, \mu}
    \bigbb{ \max\cc{ 0, R_f(z; \mu) -1 } }^2,
\label{eq:loss-D}
\end{equation}
where $(z, \qf)$ are summed over all their available bins.
The $\unpfec_f^{\rm fit}$ denotes the FEC obtained by \eq{eq:fec-param-D} plus DGLAP evolution, and is used in computing $R_f(z; \mu)$ in \eq{eq:loss-D}.
The $w_E$ controls the weight of the sum-rule penalty when $R_f > 1$.
The eventual fit satisfies $\max\cc{ 0, R_f(z; \mu) -1 } < 10^{-3}$ at every constrained point.
We use logarithmic residuals in the first term of \eq{eq:loss-D} to force the agreement of $\unpfec_f^{\rm fit}$ and $\unpfec_f^{\rm Pythia}$ in both the low-$\qf$ plateau and large-$\qf$ falling tail.

\begin{figure}[!htbp]
    \centering
    \includegraphics[scale=0.25]{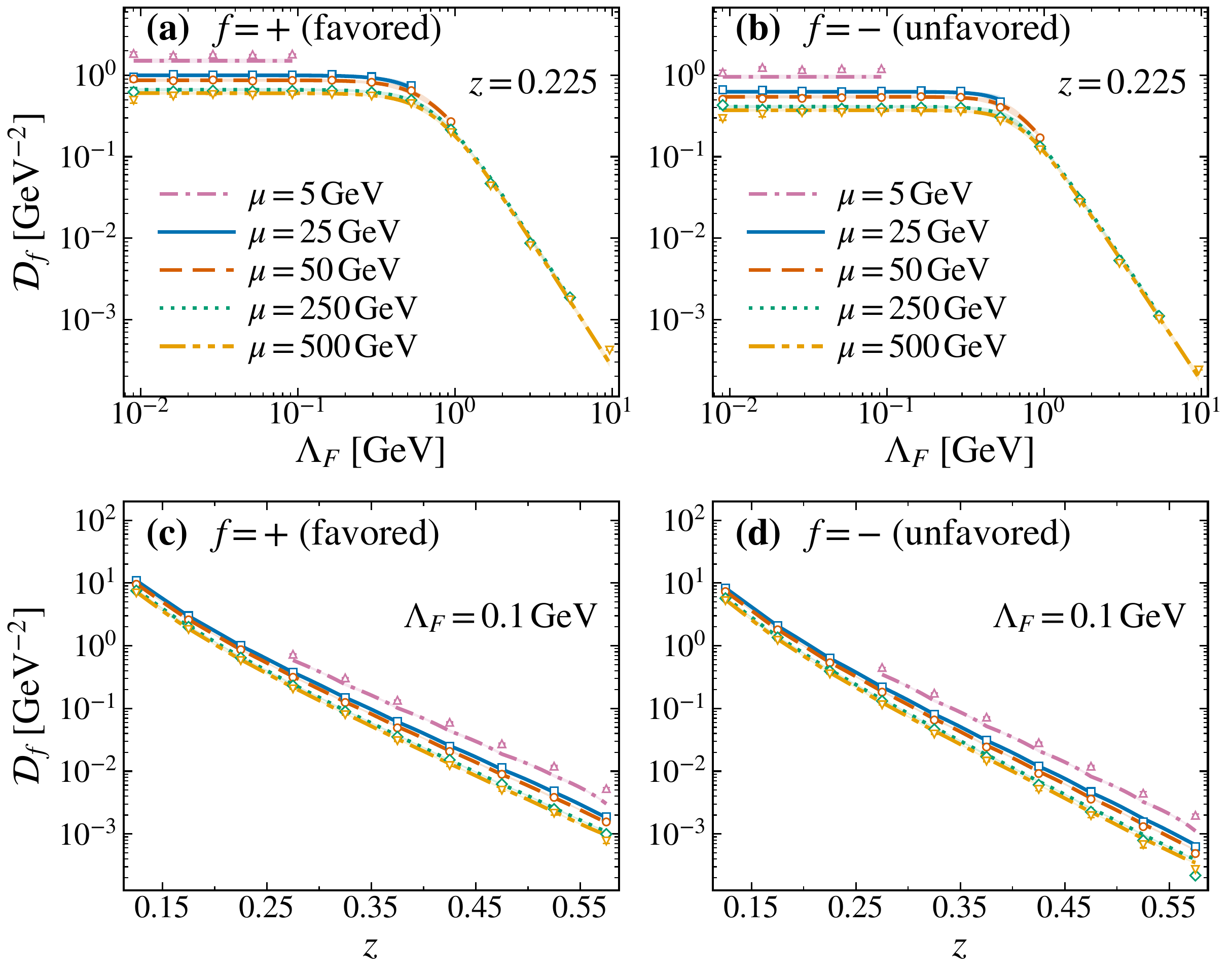}
    \caption{FECs for the flavor-favored (left) and -unfavored (right) channels at fixed $z = 0.225$ (top) and $\qf = 0.1~\GeV$ (bottom).
    The smooth curves show the fitted FECs,
    with the uncertainty bands around them coming from the fitting parameters. 
    The open markers represent the {\tt Pythia}-extracted FEC, with error bars arising from statistical uncertainties of our finite ensembles.}
\label{fig:fec-fit-D}
\end{figure}

To illustrate the fit quality, we compare the fitted FEC $\unpfec_f$ with the {\tt Pythia}-extracted one $\unpfec_f^{\rm Pythia}$ 
for the flavor-favored and -unfavored channels in \fig{fig:fec-fit-D}, 
besides the one already shown in \fig{fig:fec-dglap}(a). 
The upper row of \fig{fig:fec-fit-D} displays the $\qf^2$ dependence at fixed $z=0.225$, 
which resolves the transition from low to high $\qf$ similarly to \fig{fig:fec-dglap}(a).
The lower row shows the complementary $z$ dependence at fixed $\qf = 0.1~\GeV$. 
These comparisons show that a common set of input FECs, evolved with LO DGLAP, 
captures the main physical features implemented in {\tt Pythia}.

\begin{table}[t]
    \centering
    \caption{Parameters of the unpolarized FECs $\unpfec_f$ in \eq{eq:fec-param-D}. For $\gamma_{\unpfec, f}$ and $\kappa_{\unpfec, f}$, the means, standard deviations, and ranges describe their variation across the ten fitted $z$ nodes. 
    The intervals for the $z$-independent $\lambda_{\unpfec, f}$ are given at $68\%$ confidence level.}
    \label{tab:fit-params-D}
    \renewcommand{\arraystretch}{1.12}
    \setlength{\tabcolsep}{8pt}
    \begin{tabular}{cccccc}
    \hline\hline
    channel $f$ & $\lambda_{\unpfec, f}$ [GeV], interval
      & $\langle \gamma_{\unpfec, f} \rangle
      \pm \var \gamma_{\unpfec, f}$
      & $[\gamma_{\unpfec, f}^{\min},\gamma_{\unpfec, f}^{\max}]$
      & $\langle\kappa_{\unpfec, f} \rangle \pm \var \kappa_{\unpfec, f}$
      & $[\kappa_{\unpfec, f}^{\min}, \kappa_{\unpfec, f}^{\max}]$ \\
    \hline
    $+$ & $2.98\ [2.85,3.11]$ & $0.97\pm0.14$ & $[0.72,1.14]$ & $3.61\pm0.75$ & $[2.85,5.22]$ \\
    $-$ & $2.81\ [2.70,2.93]$ & $0.94\pm0.18$ & $[0.59,1.13]$ & $4.29\pm1.03$ & $[3.25,6.52]$ \\
    $\pi^+ / s$ & $2.34\ [2.26,2.42]$ & $0.96\pm0.52$ & $[0.41,2.31]$ & $6.39\pm4.31$ & $[0.93,12.00]$ \\
    $\pi^- / s$ & $2.28\ [2.20,2.36]$ & $0.93\pm0.48$ & $[0.47,2.21]$ & $4.70\pm2.00$ & $[1.04,7.95]$ \\
    $\pi^+ / c$ & $3.26\ [3.10,3.43]$ & $1.22\pm0.19$ & $[0.85,1.45]$ & $3.11\pm0.64$ & $[2.36,4.29]$ \\
    $\pi^- / c$ & $3.06\ [2.90,3.22]$ & $1.22\pm0.32$ & $[0.75,1.90]$ & $3.04\pm0.79$ & $[1.77,4.23]$ \\
    $\pi^+ / b$ & $6.74\ [6.09,7.47]$ & $1.95\pm0.32$ & $[1.26,2.28]$ & $2.19\pm0.17$ & $[1.94,2.55]$ \\
    $\pi^- / b$ & $8.10\ [7.26,9.04]$ & $2.24\pm0.47$ & $[1.44,2.79]$ & $2.40\pm0.49$ & $[1.90,3.59]$ \\
    $g$ & $7.57\ [5.69,10.06]$ & $4.06\pm5.55$ & $[1.03,18.00]$ & $1.63\pm0.52$ & $[1.02,2.40]$ \\
    \hline\hline
    \end{tabular}
\end{table}

The fitted parameters are given in Table~\ref{tab:fit-params-D}.
All the anomalous dimensions $\gamma_{\unpfec, f}$ are substantially larger than 0.
The average $\kappa_{\unpfec, f}$ also exceeds $1$, 
which indicates that the FECs (at least those extracted from {\tt Pythia}) tend to have sharp transitions between the perturbative region and the nonperturbative region of $\qf$.
The characteristic transition scales $\lambda_{\unpfec, f}$ exhibit clear flavor dependence.
They have similar magnitude for the flavor-favored and -unfavored channels around $2.9~\GeV$, modulo their mild $z$ dependence.

\begin{figure}[!htbp]
  \centering
  \includegraphics[scale=0.25]{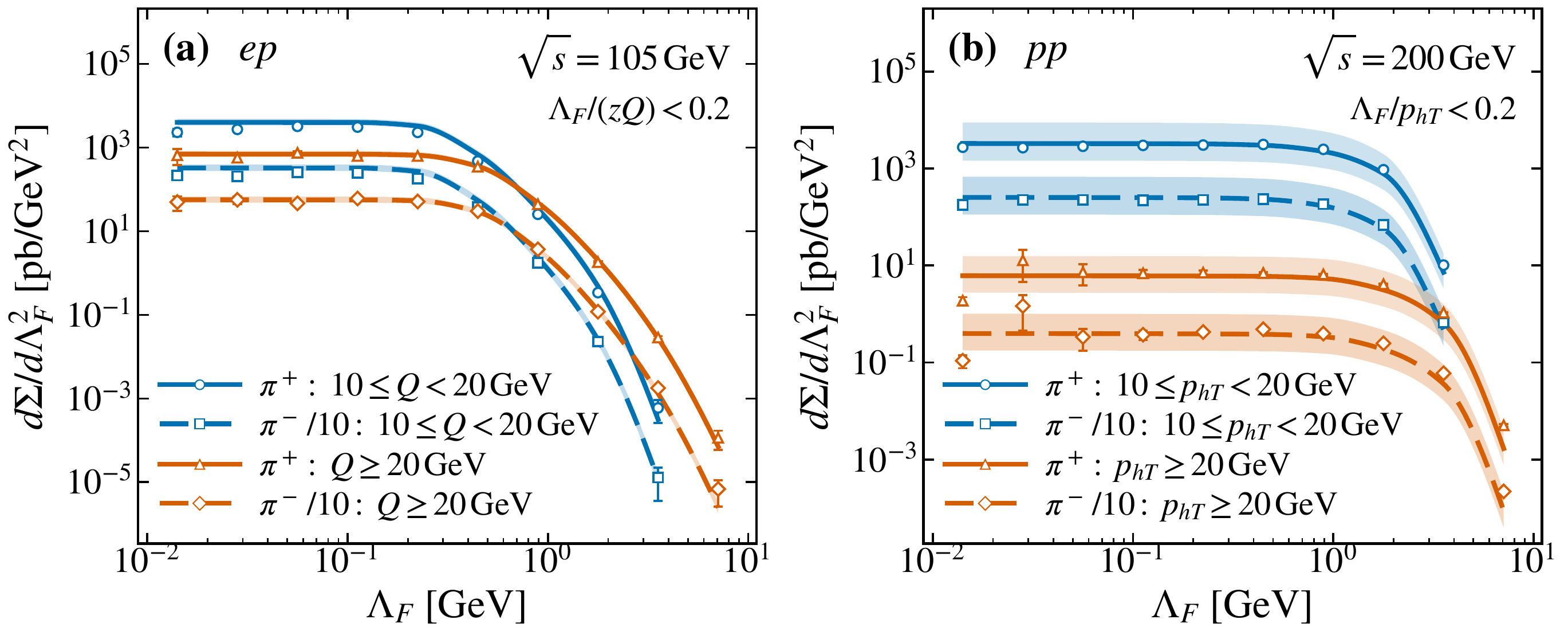}  
  \caption{Comparison of energy flow-weighted inclusive $\pi^{\pm}$ production cross sections
  between {\tt Pythia} simulations (open markers) 
  and the factorization predictions (smooth curves).
  Two kinematic bins are shown in both (a) $ep$ and (b) $pp$ collisions,
  where $\qf / (z Q) < 0.2$ and $\qf / p_{hT} < 0.2$ are required, respectively, for the validity of factorization.
  All $\pi^-$ results are divided by 10 to be distinguished from $\pi^+$ in log scale.}
\label{fig:fec-universality}
\end{figure}

With $\unpfec_f$ obtained from the $e^+e^-$ annihilation, we can apply them to the factorization formulas of $U_{ep}$ and $U_{pp}$ in \eqs{eq:ep-fact}{eq:pp-fact}, respectively,
to make predictions for the energy flow-weighted unpolarized $ep$ and $pp$ cross sections in \eqs{eq:ep-xsec}{eq:pp-xsec}.
The results are shown in \fig{fig:fec-universality} at the EIC and RHIC energies, respectively,
for both $\pi^{\pm}$ channels with two bins of $Q$ or $p_{hT}$ for each.
They are compared to the same quantities constructed from direct {\tt Pythia} simulations of these events
and exhibit good agreement across all the processes.
This constitutes a nontrivial universality check of the unpolarized FEC.

\section{Hard coefficients for $pp$ collisions in Eq.~(12)}
\label{app:pp-hard-coefs}
For the process 
$p^\uparrow(p_1) + p(p_2) \to \bb{h(p_h) + \E} + X$, 
we follow the kinematics defined in the main text and describe the events using the final-state hadron's transverse momentum $p_{hT}$, rapidity $y_h$, and azimuth $\phi_h$. 
At LO, the hard coefficients in Eq.~(12) are given by the partonic scattering,
\beq[eq:pp-partonic]
    a(k_1 = x_1 p_1) + b(k_2 = x_2 p_2) 
    \to c(k_3 = p_h / z) + d(k_4).
\eeq
The incoming partons $a$ and $b$ carry momentum fractions $x_1$ and $x_2$ of the colliding protons $p^\uparrow(p_1)$ and $p(p_2)$, respectively. 
The final-state parton $c$ fragments into the observed hadron $h$ and carries a momentum $p_h / z$ along the same direction.
The recoiling parton $d$ is unobserved and inclusively summed over, with its phase space integrated and spin summed.

The partonic Mandelstam invariants are given by
\begin{equation}
    \hat s \equiv (k_1 + k_2)^2 = x_1 x_2 s,\quad
    \hat t \equiv (k_1 - k_3)^2 = -\frac{\hat s}{2}(1-\tanh\hat y),\quad
    \hat u \equiv (k_2 - k_3)^2 = -\frac{\hat s}{2}(1+\tanh\hat y),
\label{eq:app-pp-mandelstam}
\end{equation}
where in the second step, we have given their relation to $s$ and  expressed $\hat t$ and $\hat u$ by the rapidity of the fragmenting parton $c$ in the partonic center-of-mass frame,
$$
\hat y = y_h - \frac{1}{2} \ln\pp{ \frac{x_1}{x_2} }.
$$
These invariants satisfy $\hat s+\hat t+\hat u=0$. 
The $\hat s$ sets the overall scale, 
and $\hat y$ fixes the dimensionless ratios $\hat t/\hat s$ and $\hat u/\hat s$.
The hard coefficients contain an unpolarized set $C_{abc}$ and a polarized set $T_{abc}$.
The latter applies only to the case with both $a$ and $c$ being quarks or antiquarks, where $a$ inherits a transverse spin from the proton $p^\uparrow(p_1)$
and transfers it to $c$ through the transverse spin-transfer coefficient $T_{abc}$.

At LO, the flavor of $d$ is uniquely fixed once the flavors of $(a, b, c)$ are specified, 
so coefficients $(C_{abc}, T_{abc})$ (which have summed over $d$) can be also trivially denoted by $(C_{ab\to cd}, T_{ab\to cd})$ (which have $d$ explicit).

We normalize these coefficients via the partonic cross section,
\beq
    \frac{d\hat\sigma_{ab\to cd}}{d\hat t}
    =\frac{\pi\alpha_s^2}{\hat s^2}C_{abc},\qquad
    \frac{d\Delta\hat\sigma_{a^\uparrow b\to c^\uparrow d}}{d\hat t}
    =\frac{\pi\alpha_s^2}{\hat s^2}T_{abc},
\label{eq:app-renorm-hard-coeff}
\eeq
where $\hat\sigma_{ab\to cd}$ is the unpolarized cross section, 
while $\Delta\hat\sigma_{a^\uparrow b\to c^\uparrow d}$ denotes the polarized cross section,
$$
    \frac{d\Delta\hat\sigma_{a^\uparrow b\to c^\uparrow d}}{d\hat t}
    \equiv
    \frac{d\hat\sigma_{a^\uparrow b\to c^\uparrow d}}{d\hat t} - \frac{d\hat\sigma_{a^\downarrow b\to c^\uparrow d}}{d\hat t},
$$
where on the right-hand side, both $a$ and $c$ are $100\%$ transversely polarized along (or perpendicular to) the scattering plane, and the difference is taken with the spin of $a$ flipped.
The nonzero hard coefficients are then given by 
\begin{align}
    C_{qq'q} &= C_{q \bar q' q} 
    =\frac{C_F}{N_c}\frac{\hat s^2+\hat u^2}{\hat t^2},
    \nn\\
    C_{qq'q'} &= C_{q \bar q' \bar q'}
    = \frac{C_F}{N_c}\frac{\hat s^2+\hat t^2}{\hat u^2},
    \nn\\
    C_{qqq}
    &=\frac{C_F}{N_c}\bigg[
    \frac{\hat s^2+\hat u^2}{\hat t^2}
    +\frac{\hat s^2+\hat t^2}{\hat u^2}
    -\frac{2\hat s^2}{N_c\hat t\hat u}
    \bigg],
    \nn\\
    C_{q \bar q q}
    &=\frac{C_F}{N_c}\bigg[
    \frac{\hat s^2+\hat u^2}{\hat t^2}
    +\frac{\hat t^2+\hat u^2}{\hat s^2}
    -\frac{2\hat u^2}{N_c\hat s\hat t}
    \bigg],
    \nn\\
    C_{q \bar q \bar q} 
    &=\frac{C_F}{N_c}\bigg[
    \frac{\hat s^2+\hat t^2}{\hat u^2}
    +\frac{\hat t^2+\hat u^2}{\hat s^2}
    -\frac{2\hat t^2}{N_c\hat s\hat u}
    \bigg],
    \nn\\
    C_{q \bar q q'} &= C_{q \bar q \bar q'}
    =\frac{C_F}{N_c}
    \frac{\hat t^2+\hat u^2}{\hat s^2},
    \nn\\
    C_{q\bar q g}
    &=\frac{2C_F^2}{N_c}
    \frac{\hat t^2+\hat u^2}{\hat t\hat u}
    -2C_F
    \frac{\hat t^2+\hat u^2}{\hat s^2},
    \nn\\
    C_{qgq}
    &=C_{gqg}
    = -\frac{C_F}{N_c}
    \frac{\hat s^2+\hat u^2}{\hat s\hat u}
    +\frac{\hat s^2+\hat u^2}{\hat t^2},
    \nn\\
    C_{qgg}
    &= C_{gqq}
    =-\frac{C_F}{N_c}
    \frac{\hat s^2+\hat t^2}{\hat s\hat t}
    +\frac{\hat s^2+\hat t^2}{\hat u^2},
    \nn\\
    C_{ggq} &=C_{gg\bar q}
    =\frac{T_F}{N_c}
    \frac{\hat t^2+\hat u^2}{\hat t\hat u}
    -\frac{T_F}{C_F}
    \frac{\hat t^2+\hat u^2}{\hat s^2},
    \nn\\
    C_{ggg}
    &=\frac{2N_c}{C_F}\left(
    3-\frac{\hat t\hat u}{\hat s^2}
    -\frac{\hat s\hat u}{\hat t^2}
    -\frac{\hat s\hat t}{\hat u^2}
    \right),
\label{eq:app-pp-C-table}
\end{align}
for the unpolarized ones,
and 
\begin{align}
    T_{qq'q}
    &= T_{q \bar q' q}
    = -\frac{2C_F\hat s\hat u}{N_c\hat t^2},
    \nn\\
    T_{qqq}
    &=\frac{2C_F}{N_c}\bb{-\frac{\hat s\hat u}{\hat t^2} + \frac{1}{N_c}\frac{\hat s}{\hat t}},
    \nn\\
    T_{q\bar q q}
    &=\frac{2C_F}{N_c}\bb{-\frac{\hat s\hat u}{\hat t^2} + \frac{1}{N_c}\frac{\hat u}{\hat t}},
    \nn\\
    T_{q\bar q \bar q}
    &=-\frac{2C_F}{N_c^2},
    \nn\\
    T_{qgq}
    &=\frac{2C_F}{N_c} - 2\frac{\hat s\hat u}{\hat t^2},
\label{eq:app-pp-T-table}
\end{align}
for the polarized ones~\cite{Bomhof:2006ra,Yuan:2007nd}. 
Here, both $q$ and $q'$ run through quark flavors with $q' \neq q$, and the omitted $C_{abc}$ and $T_{abc}$ coefficients with $a = q$ replaced by $a = \bar q$ can be simply obtained by charge conjugation symmetry. All other coefficients vanish.

\FloatBarrier

\end{document}